\documentclass[preprint,nofootinbib,amsmath,amssymb,amsfonts,aps,prd,eqsecnum]{revtex4-1}

\usepackage{hyperref}
\usepackage{amsthm}
\usepackage{mathrsfs}

\begin{document}

\title{Newtonian and Relativistic Cosmologies}

\author{Stephen R. Green}
\email{srgreen@uchicago.edu}
\author{Robert M. Wald}
\email{rmwa@uchicago.edu}
\affiliation{Enrico Fermi Institute and Department of Physics \\
  University of Chicago \\
  5620 S. Ellis Ave., Chicago, IL 60637, U.S.A.}

\date{\today}

\begin{abstract}

  Cosmological $N$-body simulations are now being performed using
  Newtonian gravity on scales larger than the Hubble radius. It is
  well known that a uniformly expanding, homogeneous ball of dust in
  Newtonian gravity satisfies the same equations as arise in
  relativistic FLRW cosmology, and it also is known that a
  correspondence between Newtonian and relativistic dust cosmologies
  continues to hold in linearized perturbation theory in the
  marginally bound/spatially flat case. Nevertheless, it is far from
  obvious that Newtonian gravity can provide a good {\em global}
  description of an inhomogeneous cosmology when there is significant
  nonlinear dynamical behavior at small scales.  We investigate this
  issue in the light of a perturbative framework that we have recently
  developed \cite{Green:2010qy}, which allows for such nonlinearity
  at small scales. We propose a relatively straightforward
  ``dictionary''---which is exact at the linearized level---that maps
  Newtonian dust cosmologies into general relativistic dust
  cosmologies, and we use our ``ordering scheme'' to determine the
  degree to which the resulting metric and matter distribution solve
  Einstein's equation. We find that, within our ordering scheme,
  Einstein's equation fails to hold at ``order 1'' at small scales and
  at ``order $\epsilon$'' at large scales.  We then find the additional
  corrections to the metric and matter distribution needed to satisfy
  Einstein's equation to these orders. While these corrections are of
  some interest in their own right, our main purpose in calculating
  them is that their smallness should provide a criterion for the
  validity of the original ``dictionary'' (as well as simplified
  versions of this dictionary). We expect that, in realistic Newtonian
  cosmologies, these additional corrections will be very small; if so,
  this should provide strong justification for the use of Newtonian
  simulations to describe relativistic cosmologies, even on scales
  larger than the Hubble radius.
  
\end{abstract}

\maketitle

\section{Introduction}

With the improvements in computational abilities that have taken place
in recent years, it is now feasible to do numerical simulations of
structure formation in cosmology on scales comparable to---or even
larger than---the Hubble radius. Such simulations are being carried
out by a number of groups
\cite{Evrard:2001hu,Warren:2005ey,Teyssier:2008zd,Crocce:2009mg}. However,
these simulations are being carried out using Newtonian
gravity. Although it would appear reasonable to expect Newtonian
gravity to yield a good description of phenomena
on scales much smaller than the Hubble radius---except, of course, in
the immediate vicinity of strong field objects---at first thought, it
might seem absurd that it could be expected to yield a reliable
description of phenomena on scales comparable to, or larger than, the
Hubble radius. After all, Newtonian gravity posits forces that act
instantaneously over arbitrarily large distances, whereas the
dynamical evolution laws of general relativity assert that all
influences propagate causally and that the distribution of matter
outside of one's past light cone is irrelevant. Similarly, the
Newtonian gravity description of the Hubble expansion involves
relative motion of bodies, whereas the general relativistic
description involves the expansion of space. Why should Newtonian
gravity give an accurate description of behavior on scales comparable
to---or greater than---the Hubble radius, when the relative velocity
of bodies is comparable to---or greater than---the speed of light?

Nevertheless, as we shall review in the next section, it is well known
(see, e.g., \cite{Peebles1980}) that under the assumptions of spatial
homogeneity and isotropy, the equations for a uniformly expanding
pressureless fluid (``dust'') in Newtonian gravity are identical to
the dynamical equations for a dust filled
Friedmann-Lema\^itre-Robinson-Walker (FLRW) universe in general
relativity---even in the case of nonvanishing spatial curvature. An
explanation for this remarkable correspondence can be found from the
fact that in both Newtonian gravity and general relativity, in the
presence of spherical symmetry, the behavior of a co-moving ball of
dust does not depend upon the distribution of matter outside of the
ball\footnote{This fact is closely related to the fact that there is
  no gravitational field/curvature inside a spherical shell of matter
  in Newtonian gravity (by Newton's theorem) and general relativity
  (by Birkhoff's theorem).}. Thus, in both Newtonian gravity and
general relativity, the dynamical behavior of a co-moving ball of dust
in a homogeneous, isotropic universe is the same as it would be if
that ball were placed in an empty, asymptotically flat
spacetime. However, for a sufficiently small ball of dust in an
otherwise empty spacetime, Newtonian gravity should be an excellent
approximation to general relativity. Thus, for a sufficiently small
ball, the density and co-moving radius of the ball must satisfy the
same dynamical equations in Newtonian gravity and general
relativity. This implies that the equations for the density and
comoving radius in a homogeneous, isotropic Newtonian dust cosmology
must coincide with the equations for the density and scale factor in a
FLRW dust cosmology. This correspondence continues to hold in the
presence of a cosmological constant term in the Newtonian and general
relativistic equations.

The above argument relies crucially on exact spherical symmetry. Thus,
one might expect that no such correspondence between Newtonian and
relativistic cosmologies would hold if one perturbs the homogeneous,
isotropic solutions away from spherical symmetry. Remarkably, however,
the correspondence between Newtonian and relativistic cosmologies
extends into the regime of linearized perturbation theory in the case
of perturbations off of a spatially flat FLRW dust cosmology. More
precisely, as pointed out by Bardeen \cite{Bardeen:1980kt} and will be
reviewed in the next section, the scalar gauge-invariant variables of
linearized relativistic perturbation theory obey exactly the same
equations as the variables describing linearized irrotational dust
perturbations of the corresponding Newtonian cosmology. Furthermore,
it is not difficult to see that this correspondence extends to the
vector case as well, i.e., vector gauge-invariant variables of
linearized relativistic perturbation theory obey exactly the same
equations as the corresponding Newtonian variables describing
vorticity perturbations. Thus, the scalar and vector
sectors\footnote{The tensor modes correspond to additional degrees of
  freedom present only in general relativity, and they have no
  Newtonian correspondence.} of linearized relativistic perturbation
theory off of a spatially flat FLRW dust model are in exact
correspondence with arbitrary Newtonian perturbations off of the
corresponding Newtonian dust cosmology.

Further justification for the validity of the Newtonian approximation
in cosmology is provided by the work of Oliynyk
\cite{Oliynyk:2009bc,Oliynyk:2009qa} (see also Futamase
\cite{Futamase:1996fk}).  Oliynyk rigorously proved that for a given
$3$-torus Newtonian cosmology, there exists a one-parameter family of
general relativistic solutions that limits to this Newtonian
cosmology, thus showing that there are general relativistic solutions
that are arbitrarily close to the Newtonian solution.  However, for
Oliynyk's one-parameter families, the ratio of the size of the
universe to the Hubble radius goes to zero in the
limit\footnote{Oliynyk formulated his limit as one in which the
  $3$-torus remains of fixed size, but the speed of light---and the
  Hubble radius---goes to infinity. By re-scaling the spatial
  coordinates, his limit can be reformulated as one in which the speed
  of light remains constant and the Hubble radius goes to a well
  defined limit, but the size of the $3$-torus then approaches zero in
  the limit.}. Thus, the general relativistic solutions proven by
Oliynyk to be very close to a Newtonian solution have size small
compared with the Hubble radius, and thus have no ``long wavelength
part''.  Thus, Oliynyk's results do not directly address the issue of
whether Newtonian simulations on scales comparable to the Hubble
radius correspond closely to a general relativistic solution, but it
can be viewed as providing additional justification for the validity
of Newtonian gravity on scales small compared with the Hubble radius.

Taken together, the above considerations strongly suggest that for a
universe that is sufficiently close to a spatially flat FLRW dust
model, Newtonian gravity should provide a good description of
structure formation on all scales. However, the situation is far from
straightforward for the following reasons: (i) Although, as described
above, there is a correspondence at linearized order between Newtonian
theory and general relativity, the ``dictionary'' needed to translate
a linearized Newtonian solution into metric and matter perturbations
in any particular gauge is nontrivial, and it is not obvious how this
dictionary compares with standard dictionaries used for the Newtonian
and post-Newtonian approximations to general relativity on small
scales. Thus, it is not obvious how to produce a ``global dictionary''
that works on all scales. (ii) If one has a candidate global
dictionary, it is not obvious how to formulate criteria to determine
whether the resulting general relativistic spacetime is ``sufficiently
close'' to a solution to Einstein's equation to trust its
predictions. The main complication here is that the failure to take
post-Newtonian corrections into account on small scales will cause the
general relativistic spacetime to fail to satisfy Einstein's equation
by a larger amount than the failure to properly account in any way for
the long wavelength perturbations. For most applications in cosmology,
the tiny post-Newtonian corrections to the metric and matter motion on
small scales are of no interest, but the leading order deviation of
the metric and matter density from a FLRW model on large scales is of
great interest. Thus, the proper criteria for being ``sufficiently
close'' to a solution to Einstein's equation must take into account
the distinction between small scales and large scales. (iii) One would
like to know explicitly what the dominant corrections to Newtonian
cosmology are, both to be able to quantitatively judge its reliability
and to be able to make its predictions more accurate.

The difficulties in addressing the above issues stem from the fact
that the approximations of Newtonian gravity (which, {\it a priori},
is expected to be good on small scales) and linearized perturbation
theory (which, {\it a priori}, is expected to be good on large scales)
are incompatible. Specifically, in the Newtonian gravity approximation
certain nonlinear terms in the equations are kept (as they must be at
small scales), but it is essential that time derivatives of quantities
be small compared with space derivatives
\cite{Oliynyk:2009qa,Oliynyk:2009bc,Futamase:1983,lrr-2007-2}. By
contrast, linearized perturbation theory allows time derivatives of
quantities to be comparable to their space derivatives (as they must
be at large scales), but it is essential that all nonlinear terms be
negligible. In order to properly treat phenomena on all scales, one
needs an approximation scheme that can accommodate nonlinear phenomena
on small scales but treats time derivatives on the same footing as
space derivatives on large scales.  We recently proposed an approach
that accomplishes this \cite{Green:2010qy}, and we will apply this
approach here\footnote{Our approach is closely related to
  \cite{Baumann:2010tm}; see also
  \cite{Carbone:2004iv}.}.

The main questions we wish to address in this paper can now be stated
concretely as follows.  Suppose that a Newtonian cosmological
simulation has been performed on a $3$-torus (i.e., periodic boundary
conditions), where the size of the $3$-torus may be larger than the
Hubble radius. For convenience, we assume that the Newtonian solution
has been presented as a continuum solution---i.e., that suitable
smoothing has been done if the solution was produced from an $N$-body
simulation.  We would like to know the following: (1) What general
relativistic spacetime and dust matter distribution should we
associate to this Newtonian cosmology, i.e., what ``global
dictionary'' should we use?  (2) To what extent is this spacetime a
solution to Einstein's equation, i.e., what are the leading order
terms in Einstein's equation that fail to be satisfied? (3) What are
the leading order corrections to the metric and dust distribution that
improve the accuracy of this solution, and how large are these
corrections?

Our approach will be to use the framework of \cite{Green:2010qy} to
provide a ``counting scheme'' for the sizes of terms in Einstein's
equation.  We will start with a candidate ``global dictionary,''
which is suggested by the known correspondences between Newtonian
gravity and general relativity in the exactly homogeneous and
isotropic case and at the linearized level. We will then see that in
our counting scheme, the resulting general relativistic spacetime
fails to satisfy Einstein's equation to $O(1)$ at small scales and to
$O(\epsilon)$ at large scales. The main effort in our paper will then
be to find the corrections to the metric and dust distribution that,
within our counting scheme, improve the accuracy of the solution to
$O(1)$ at small scales and to $O(\epsilon)$ at large scales.  It should
be emphasized that we shall not {\it prove} existence of a
one-parameter family of solutions to Einstein's equation with the
properties we desire---a far more difficult task than solving for
leading order corrections. Nevertheless, if the leading order
corrections we obtain are small compared with terms appearing in the
original global dictionary, we believe that this provides a strong
indication that there is a general relativistic solution that
corresponds closely to the Newtonian cosmology. Conversely, if these
corrections are not negligibly small, then either the Newtonian
cosmology is not providing a sufficiently accurate representation of
the general relativistic spacetime or the dictionary being used will
have to be significantly modified.

Our analysis also addresses concerns that have been expressed with
regard to the use of a ``Newtonianly perturbed FLRW metric,'' which
corresponds to the using the ``abridged dictionary'' given by
\eqref{eq:adict-L1}--\eqref{eq:adict-L3} below.  Ishibashi and Wald
\cite{Ishibashi:2005sj} have argued that this metric should provide an
excellent description of our universe.  However, several authors
\cite{vanElst:1998kb,Buchert:2009wj} have objected to the use of this
metric on the grounds that, if taken literally, and dust peculiar
velocities are ignored, then strong constraints relating to exact
solutions of Einstein's equation apply, and the metric is only able to
describe a spatially homogeneous continuum.  Other concerns have been
raised by Rasanen \cite{Rasanen:2010wz}.  The spacetime metric and
dust matter distribution that we produce in this paper---as summarized
in section \ref{sec:summary}---solves Einstein's equation to a much
higher degree of accuracy than the Newtonianly perturbed FLRW metric
does, and, in particular, fully takes into account peculiar velocities
and leading nonlinear terms in the Einstein equation.  No
inconsistencies of any kind are encountered in obtaining this much
more accurate solution. Thus, the approximate solution considered in
\cite{Ishibashi:2005sj} should be fully justified provided only that
the corrections to \eqref{eq:adict-L1}--\eqref{eq:adict-L3} given in
section \ref{sec:summary} are negligibly small, as we argue is the
case.

We remark that if one has an equation ${\mathcal E} (F) = 0$, one must
draw a clear distinction between having a quantity $f$ that
approximately solves this equation (i.e., ${\mathcal E} (f) \approx
0$) as compared with having a quantity $f$ that is an approximate
solution (i.e., $f \approx F$ for some exact solution ${\mathcal E}
(F) = 0$). If the equation is suitably well posed, if ${\mathcal E}
(f) \approx 0$, and if $F$ is the exact solution with the same initial
data as $f$, then $f$ and $F$ will remain close to each other for
sufficiently early times. However, $f$ may fail to remain close to $F$
at late times because of the build-up of secular effects. For example,
the Newtonian solution for the motion of Mercury solves the general
relativistic equations of motion to an excellent approximation at all
times, but provides a very poor approximation to the general
relativistic solution for the position of Mercury after $\sim 10^6$
years. In this paper, we are concerned with the issue of obtaining
general relativistic spacetimes that solve Einstein's equation to an
excellent approximation at all times, but we will not be concerned
with the issue of whether these spacetimes provide good global-in-time
approximations to exact solutions of Einstein's equation.

In the next section, we shall review the correspondence between
homogeneous, isotropic Newtonian cosmology with dust matter and FLRW
models in general relativity, as well as the correspondence at the
linearized level between these models in the marginally
bound/spatially flat case.  On the basis of this correspondence, we
will propose a dictionary \eqref{eq:dict-L1}--\eqref{eq:dict-L5} to
translate Newtonian cosmologies into general relativistic
spacetimes. In section \ref{sec:improveddictionary}, we will apply our
counting scheme \cite{Green:2010qy} to analyze how well Einstein's
equation is being satisfied, and we will obtain corrections needed to
satisfy Einstein's equation to $O(1)$.  We will then obtain the
further modifications to the metric and dust distribution needed to
obtain a solution to Einstein's equation to $O(\epsilon)$ at large
scales. These corrections are of some interest in their own right.
For example, as we shall see in Appendix \ref{sec:modifiedbackground},
there are small modifications of some global properties of the
cosmology, such as a slight modification of the expansion rate and the
introduction an (even smaller) anisotropic expansion.  However, our
main purpose in determining these corrections is to provide a
criterion for the validity of the Newtonian cosmology as translated
into a general relativistic spacetime via the dictionary
\eqref{eq:dict-L1}--\eqref{eq:dict-L5} and/or its abridgment
\eqref{eq:adict-L1}--\eqref{eq:adict-L3} or its simplification
\eqref{eq:sdict-L1}--\eqref{eq:sdict-L3}: The full set of metric and
matter corrections to our original dictionary are given by
eqs.~\eqref{eqn:d1B}--\eqref{eqn:d14B} and these corrections can be
computed straightforwardly for any given Newtonian cosmology.  The
smallness of these corrections should provide a reliable criterion for
judging the validity of using a Newtonian simulation with the
dictionary \eqref{eq:dict-L1}--\eqref{eq:dict-L5} (or its abridgment
or simplification) to describe a relativistic cosmology.

\section{Background and Linearized Correspondence}

In this section we shall review the correspondence between
homogeneous, isotropic Newtonian dust cosmology and FLRW models, as
well as the correspondence between linearized perturbations of these
models. We shall then propose a dictionary---valid to linearized
order---the accuracy of which will be evaluated and improved upon in
the following section.

In Newtonian gravity, the gravitational field is described by a
Newtonian potential $\phi$, and the dust matter is described by a mass
density $\rho$ and a velocity field $v^i$.  The Newtonian field
potential is related to the mass density by the Poisson equation
\begin{equation}\label{eq:NG}
  \partial^i\partial_i\phi + \Lambda = 4\pi\rho\,,
\end{equation}
which we have generalized to allow for the presence of a cosmological
constant $\Lambda$.  In addition, the matter variables must satisfy
mass conservation and Euler equations, which, for dust matter, take
the form
\begin{eqnarray}\label{eq:Nmass}
  \partial_t \rho + \partial_i(\rho v^i) &=& 0\,,\\
  \label{eq:NEuler}\partial_t (\rho v^i) + \partial_j(\rho v^i v^j) &=& -\rho\partial^i \phi\,.
\end{eqnarray}
In these equations, the flat Euclidean metric of space is used to contract
indices.

\subsection{Background Correspondence}

As a cosmological ansatz, we seek a solution to the above equations of Newtonian gravity
in which the density is spatially uniform, 
$\rho=\rho_0(t)$, and the velocity field is uniformly expanding, 
$v^i=H(t) {x}^i$. Note that $H$ is related to the radius, $a$, of any comoving ball by
\begin{equation}
H=\frac{1}{a}\frac{da}{dt} \, .
\end{equation}
Since $\partial_iv^i\equiv 3H(t)$, \eqref{eq:Nmass} implies that
\begin{equation}
  \partial_t \rho_0 + 3 H \rho_0 = 0\,,
\end{equation}
from which it follows that 
\begin{equation}
\rho_0=\rho_{0,\text{init}} a^{-3} \, .
\end{equation}
Using mass conservation, we can eliminate $\rho$ from the Euler equation,
\begin{equation}
  \partial_tv^i+v^j\partial_jv^i=-\partial^i\phi\,.
\end{equation}
The Poisson equation, \eqref{eq:NG}, has the non-singular
solution
\begin{equation}
  \phi_0 = \frac{2\pi}{3}\rho_0r^2-\frac{\Lambda}{6}r^2+A(t)\,.
\end{equation}
Substituting for $v$ and $\phi$, we obtain
\begin{equation}
  \frac{dH}{dt}+H^2=-\frac{4\pi}{3}\rho_0+\frac{\Lambda}{3}\,,
\label{F1}
\end{equation}
which is one of the Friedmann equations.
To obtain the other Friedmann equation, we rewrite this equation as
\begin{equation}
  \frac{1}{a}\frac{d^2a}{dt^2}=-\frac{4\pi}{3}\frac{\rho_{0,\text{init}}}{a^3}+\frac{\Lambda}{3}\,.
\end{equation}
Integrating once, we obtain
\begin{equation}
  H^2=\frac{8\pi}{3}\rho_0+\frac{\Lambda}{3}-\frac{k}{a^2}\,,
\label{F2}
\end{equation}
where $k$ is a constant of integration. By choosing the size of the
comoving ball appropriately, we may choose $k$ to take the values $0,
\pm 1$. When $\Lambda = 0$, the value of $k$ determines whether the
universe is unbound and expands forever ($k=-1$), is marginally bound
and expands forever but with expansion velocity approaching zero ($k =
0$), or is bound and will recollapse within finite time ($k=+1$). Of
course, in Newtonian gravity, $k$ does not have any interpretation in
terms of spatial curvature; space is always Euclidean.

Equations \eqref{F1} and \eqref{F2} are precisely the equations
satisfied by dust FLRW models in general relativity. The underlying
reason for this exact correspondence was discussed in the
Introduction.

\subsection{Linearized Correspondence}

We first re-write the exact Newtonian equations relative to some
(arbitrarily chosen) background solution of the previous section. We
introduce comoving coordinates
\begin{equation}
    x'{}^i=\frac{x^i}{a}\,.
\end{equation}
We then define the velocity variable $v'{}^i$ by
\begin{equation}
 v'{}^i\equiv a \frac{dx'{}^i}{dt'}=\frac{dx^i}{dt}-\frac{1}{a}\frac{da}{dt}x^i=v^i-Hx^i\,,
\end{equation}
so $v'{}^i$ measures the velocity relative to the Hubble flow of the background solution.
We also define density and potential deviations from the background, $\delta$ and $\psi$, by
\begin{eqnarray}
  \rho&=&\rho_0(1+\delta)\,,\\
  \phi&=&\phi_0+\psi\,.
\end{eqnarray}
In terms of these quantities, the Newtonian equations are
\begin{eqnarray}
  \partial^{i'}\partial_{i'}\psi&=&4\pi a^2\rho_0\delta\,,\\
  \partial_{t'}\delta + \frac{1}{a}\partial_{i'}\left((1+\delta) v'^{i'}\right)&=&0\,,\\
  \partial_{t'}v'^{i'}+\frac{1}{a}v'^{j'}\partial_{j'}v'^{i'}+Hv'^{i'}&=&-\frac{1}{a}\partial^{i'}\psi\,.
\end{eqnarray}
From now on, since we will always work in comoving coordinates, we shall
drop the primes.

Next, we re-write these equations using ``conformal time''
$\tau$ defined by
\begin{equation}
\frac{d\tau}{dt} =\frac{1}{a} \, .
\end{equation}
We also denote derivatives with respect
to conformal time with overdots.  In terms of the conformal time variable, 
the Newtonian background equations are
\begin{eqnarray}
  \frac{\dot{a}^2}{a^2}&=&\frac{8\pi}{3}a^2\rho_0+\frac{\Lambda}{3}a^2-k\,,\\
  \frac{d}{d\tau}\left(\frac{\dot{a}}{a}\right)&=&-\frac{4\pi}{3}a^2\rho_0+\frac{\Lambda}{3}a^2\,,\\
  \dot{\rho}_0+3\frac{\dot{a}}{a}\rho_0&=&0\,,
\end{eqnarray}
and the Newtonian equations for the quantities describing the deviations from the background are
\begin{eqnarray}
  \label{eq:NG-pert}\partial^{i}\partial_{i}\psi_N&=&4\pi a^2\rho_0\delta_N\,,\\
  \label{eq:Nmass-pert}\dot{\delta}_N + \partial_{i}\left((1+\delta_N) v_N^{i}\right)&=&0\,,\\
  \label{eq:NEuler-pert}\dot{v}_N^{i}+v_N^{j}\partial_{j}v_N^{i}+\frac{\dot{a}}{a}v_N^{i}&=&-\partial^{i}\psi_N\,,
\end{eqnarray}
where we have now added a subscript $N$
so that these Newtonian quantities can be easily distinguished from 
the corresponding general relativistic quantities that we will introduce later.  
We emphasize that
\eqref{eq:NG-pert}--\eqref{eq:NEuler-pert} are exact. We shall assume below that 
these equations are solved on a $3$-torus, i.e., a ``box'' at fixed comoving coordinates
(of the background solution) with periodic boundary conditions.

Linearizing \eqref{eq:NG-pert}--\eqref{eq:NEuler-pert}
about the background solution, we obtain
\begin{eqnarray}
  \label{eq:NG-pertL}\partial^{i}\partial_{i}\psi^{(1)}_N&=&4\pi a^2\rho_0\delta^{(1)}_N\,,\\
  \label{eq:Nmass-pertL}\dot{\delta}^{(1)}_N + \partial_{i}v^{(1)i}_N&=&0\,,\\
  \label{eq:NEuler-pertL}\dot{v}^{(1)i}_N+\frac{\dot{a}}{a}v^{(1)i}_N&=&-\partial^{i}\psi^{(1)}_N\,.
\end{eqnarray}
We now compare the linearized Newtonian equations with the linearized
general relativistic equations about a dust FLRW model.  In
\cite{Bardeen:1980kt}, Bardeen decomposed linearized metric and
stress-energy perturbations into their scalar, vector, and tensor
parts, which evolve independently. He then introduced gauge invariant
quantities to describe these perturbations. In the case of a perfect
fluid, the two scalar gauge invariant variables describing metric
perturbations are related---in his notation, $\Phi_H=-\Phi_A$---so the
scalar perturbations are fully described by $\Phi_H$ (or $\Phi_A$),
the scalar part of the velocity perturbation, $v_s^i$, and a
density perturbation variable $\epsilon_m$ (defined by
eqs.~(3.9)--(3.11) and (3.13) of \cite{Bardeen:1980kt}).  We can
similarly decompose a linearized Newtonian perturbation: $\psi_N^{(1)}$ and
$\delta_N^{(1)}$ are scalar quantities, and the velocity perturbation can be
decomposed as
\begin{equation}
v_{N}^{(1)i} = v_{N\text{s}}^{(1)i} + v_{N\text{v}}^{(1)i} \,,
\label{sv}
\end{equation}
where $v_{N\text{s}}^{(1)i} $ can be written as a gradient and $\partial_i
v_{N\text{v}}^{(1)i} = 0$.  Newtonian perturbations have no tensor part.
It is then straightforward to see that, 
as pointed out by Bardeen \cite{Bardeen:1980kt},
under the correspondence
\begin{eqnarray}
  \label{eq:dict-GI1}\psi_N^{(1)}&\longleftrightarrow&\Phi_A=-\Phi_H\,,\\
  \label{eq:dict-GI2}v_{N\text{s}}^{(1)i}&\longleftrightarrow&v_s^i\,,\\
  \label{eq:dict-GI3}\delta_N^{(1)}&\longleftrightarrow&\epsilon_m\,,
\end{eqnarray}
the linearized Newtonian equations become identical to the equations
describing scalar perturbations of a spatially flat dust cosmology as
given by eqs.~(4.3), (4.5) and (4.8) of \cite{Bardeen:1980kt}.  Note
that this correspondence holds only for perturbations of spatially
flat models, i.e., there are additional terms in the linearized
Einstein equation when the background solution has nonvanishing
spatial curvature.

It is not difficult to see that the exact correspondence between
linearized Newtonian gravity and general relativistic perturbations of
spatially flat models extends to vector perturbations as well with
\begin{equation}
  \label{eq:dict-GI4}v_{N\text{v}}^{(1)i}\longleftrightarrow v_c^i\,,
\end{equation}
where the gauge invariant quantity $v_c^i$ is defined by Bardeen's
eq.~(3.23).  Specifically, the vector part of the linearized Euler
equation \eqref{eq:NEuler-pertL} for $v_{N\text{v}}^{(1)i}$ is identical to
Bardeen's eq.~(4.13) for $v_c^i$. There are no additional Newtonian
equations for vector perturbations. However, there is an additional general
relativistic equation [Bardeen's eq.~(4.12)],
\begin{equation}\label{eq:Psi}
  \partial^j\partial_j\Psi^i=-16\pi a^2\rho_0v_c^i\,,
\end{equation}
which is a Poisson equation for a quantity $\Psi^i$ not present in Newtonian
theory. On a torus, this equation has a solution if and only if there is no spatially
homogeneous part of $v_c^i$, i.e., 
\begin{equation}
  \int d^3x\,v^{(1)i}_{N\text{v}}=0 \, .
\label{intv}
\end{equation}
This equation must hold for all for all times $\tau$, but it is easily
checked that this equation is preserved under time evolution, so it
suffices to impose it at any one time. Thus, \eqref{intv} is a
constraint that must be imposed upon a linearized Newtonian solution
in order that it correspond to a linearized solution of Einstein's
equation under the correspondence \eqref{eq:dict-GI4}

In summary, provided only that the constraint \eqref{intv} is
satisfied, there is an exact correspondence, given by
\eqref{eq:dict-GI1}--\eqref{eq:Psi}, between the complete linearized
Newtonian equations for dust matter off of a homogeneous and isotropic
background and the scalar and vector parts of the linearized Einstein
equation off of a spatially flat dust FLRW background. As previously
noted, there are no counterparts to tensor perturbations in Newtonian
gravity, i.e., general relativity has these additional degrees of
freedom not present in Newtonian gravity.

\subsection{A Proposed Dictionary}\label{sec:proposeddictionary}

Based upon the results of the previous subsections, we now shall
propose a dictionary that translates a solution $(\psi_N,
\delta_N, v_N^i)$ of the exact Newtonian equations
\eqref{eq:NG-pert}--\eqref{eq:NEuler-pert} into a general relativistic
spacetime metric $g_{ab}$ and dust matter stress-energy tensor $T_{ab}
= \rho u_a u_b$. In the next section, we shall investigate the extent
to which $(g_{ab}, T_{ab})$ satisfies Einstein's equation as well as
what further corrections need to be made to $(g_{ab}, T_{ab})$ to make
it solve Einstein's equation to higher accuracy.

First, the Newtonian equations \eqref{eq:NG-pert}--\eqref{eq:NEuler-pert}
were written relative to
a ``background solution'' of \eqref{F1} and \eqref{F2} 
with ``scale factor'' $a$ and mass density $\rho_0$. Since 
these Newtonian background equations are identical to the 
equations for a dust filled FLRW general relativistic spacetime, we
define our dictionary so that it
associates the corresponding FLRW spacetime to this background solution.
Thus, we have defined our dictionary for the case 
$\psi_N=\delta_N=v_N^i=0$. In the following, we shall restrict consideration
to the case where $k=0$ for the background solution, since this is the
only case where we expect a good dictionary to exist when deviations
from homogeneity and isotropy occur. We shall assume that 
periodic boundary conditions have been imposed on the Newtonian
background solution, so that the corresponding FLRW background solution
has $3$-torus spatial slices. For convenience, we assume that the co-moving
spatial coordinates of the Newtonian and FLRW background solutions range between
$0$ and $1$.

In order for our dictionary to produce a definite general relativistic
spacetime, we must make a choice of gauge for the metric.  In the
context of linearized perturbation theory, a natural and very useful
gauge choice is the longitudinal gauge, in which the metric takes the
form
\begin{equation}\label{eqn:Poissongauge}
  ds^2=a^2(\tau)\left[-(1+2A)d\tau^2 - 2B_idx^id\tau + \left((1+2H_L)\delta_{ij}+h_{ij}\right)dx^idx^j\right]\,,
\end{equation}
where $\partial^iB_i=0$, $\partial^jh_{ij}=0$ and
$h^i_{\phantom{i}i}=0$, and spatial indices $i,j,k,\ldots$ are raised
and lowered with the background flat Euclidean metric
$\delta_{ij}$. In the context of linearized perturbation theory, the
quantities $A$, $B_i$, $H_L$, and $h_{ij}$ represent the metric
perturbation, and it can be shown that an arbitrary metric
perturbation can be put in the form \eqref{eqn:Poissongauge} by an
infinitesimal gauge transformation.  It also can be shown that this
gauge is essentially unique, i.e., there is essentially no additional
gauge freedom that maintains the form
\eqref{eqn:Poissongauge}. However, linearized perturbation theory is
not adequate for our purposes, since our dictionary is required to map
Newtonian solutions that differ by a finite amount from the Newtonian
background solution into metrics that differ by a finite amount from
an FLRW model. Nevertheless, it should be possible to show via the
implicit function theorem that for metrics that differ from an FLRW
model by a sufficiently small but finite amount, the metric form
\eqref{eqn:Poissongauge}---with $\partial^iB_i=0$,
$\partial^jh_{ij}=0$ and $h^i_{\phantom{i}i}=0$---always can be
imposed by a (nonlinear) gauge transformation. We shall not attempt to
prove such a result here, and will merely take
\eqref{eqn:Poissongauge} as an ansatz for the metric in constructing
our dictionary. However, we believe that imposition of the metric form
\eqref{eqn:Poissongauge} does not involve any loss of generality if
the metric is sufficiently close to an FLRW model.

The stress-energy tensor of dust in the general relativistic spacetime
takes the form
\begin{equation}
  T_{ab}=\rho u_a u_b\,.
  \label{Tab}
\end{equation}
We define the three-velocity, $v^i$, of the dust to be such that 
the components, $u^\mu$, of the four-velocity in our gauge are proportional to
$(1,v^i)$.  Normalizing using the metric form \eqref{eqn:Poissongauge}, we obtain  
\begin{equation}
  u^\mu=\frac{1}{a\sqrt{1+2A+2B_jv^j-\left((1+2H_L)\delta_{jk}+h_{jk}\right)v^jv^k}}(1,v^i)\,.
\end{equation}
Thus, this equation gives the formula for the $4$-velocity $u^a$ appearing in
\eqref{Tab} in terms of the $3$-velocity $v^i$ that will be specified by our dictionary below.
We define the fractional density perturbation $\delta$ in the general relativistic model via
\begin{equation}
  \rho=\rho_0(1+\delta)\,.
  \label{delta}
\end{equation}

As already stated above, the Newtonian solution is specified by
$(\psi_N, \delta_N, v_N^i)$.  With the above gauge choice, the general
relativistic spacetime and matter distribution is specified by $(A,
B_i, H_L, h_{ij}, \delta, v^i)$. Our proposed dictionary will
therefore be defined by providing formulas for $A$, $B_i$, $H_L$,
$h_{ij}$, $\delta$, and $v^i$ in terms of the Newtonian variables. To
obtain this dictionary, we start by taking the formulas that hold at
linearized order under the correspondence of the previous section,
which we obtain by expressing the Bardeen variables appearing in
\eqref{eq:dict-GI1}--\eqref{eq:dict-GI4} in terms of $(A, B_i, H_L,
h_{ij}, \delta, v^i)$. Then we improve our definition for $B_i$ (and
correspondingly for $v^i$) by requiring consistency with the nonlinear
momentum constraint at small scales, leading to the replacement of $\rho_0 v^i$
by $\rho_0(1 + \delta_N)v^i$. We thereby propose the following
dictionary:
\begin{eqnarray}
  \label{eq:dict-L1} A=-H_L &=& \psi_N\,,\\
  \label{eq:dict-L2} (1+\delta_N)v^i&=&(1+\delta_N)(v^i_N+ B^i)-\left.\overline{(1+\delta_N)v_N^i}\right|_{\text{v}}\,,\\
  \label{eq:dict-L3}\delta &=& \delta_N-\frac{3}{4\pi\rho_0a^2}\left[\left(\frac{\dot{a}}{a}\right)^2\psi_N+\frac{\dot{a}}{a}\dot{\psi}_N\right]\,,\\
  \label{eq:dict-L4} h_{ij} &=& 0\,,
\end{eqnarray}
and $B^i$ is the solution to the equation 
\begin{equation}
\partial^j\partial_j B^i = -16\pi\rho_0a^2\left.\left((1+\delta_N)v_N^i-\overline{(1+\delta_N)v_N^i}\right)\right|_{\text{v}}\, ,
\label{eq:dict-L5}
\end{equation}
with\footnote{\label{Bbar}The condition $\overline{B}_i=0$ can be imposed by using the 
coordinate freedom $x^i \rightarrow x^i + F^i(t)$.} $\overline{B}_i=0$,
where the overbar denotes spatial average, i.e., 
\begin{equation}
  \overline{f}\equiv\int d^3x f\, .
  \label{bar}
\end{equation}
(Recall that the comoving spatial coordinates are assumed to range 
from $0$ to $1$.) In \eqref{eq:dict-L2} and \eqref{eq:dict-L5}, the notation
$|_{\text{v}}$ denotes the ``vector part''
of a quantity in a decomposition of the type \eqref{sv}.

At large scales, one would expect the vector part of $v^i$ to be
negligible because in linear perturbation
theory, vector modes are known to decay
\cite{Bardeen:1980kt}. In addition, comparing the Poisson equation for
$B^i$ with the Poisson equation for $\psi_N$, one would expect $B^i$
to be smaller than $\psi_N$ by order $v/c$ at small scales, and thus
$B^i$ should be negligible compared with $A$ and $H_L$ at all
scales. Thus, \eqref{eq:dict-L2}
should yield a negligibly small correction to the equation $v^i =
v^i_N$. Thus, under normal circumstances, it should be acceptable to
replace our proposed global dictionary with the following {\em
  abridged version} of the dictionary:
\begin{eqnarray}
  \label{eq:adict-L1} A=-H_L &=& \psi_N\,,\\
  \label{eq:adict-L2} v^i&=&v^i_N \,,\\
  \label{eq:adict-L3}\delta &=& \delta_N-\frac{3}{4\pi\rho_0a^2}\left[\left(\frac{\dot{a}}{a}\right)^2\psi_N+\frac{\dot{a}}{a}\dot{\psi}_N\right]\, ,
\end{eqnarray}
together with $B^i = h_{ij} = 0$. This abridged dictionary corresponds
to a continuum version of the dictionary\footnote{Their definition of
  $\rho$ differs from ours by a term involving the perturbed volume
  element.} given by Chisari and Zaldarriaga \cite{Chisari:2011iq}.

Finally, on small scales $\delta_N$ should dominate the other terms
appearing on the right side of \eqref{eq:dict-L3}. Thus, on scales
much smaller than the Hubble radius, it should be possible to use the
following {\em simplified version} of the dictionary:
\begin{eqnarray}
  \label{eq:sdict-L1} A=-H_L &=& \psi_N\,,\\
  \label{eq:sdict-L2} v^i&=&v^i_N \,,\\
  \label{eq:sdict-L3}\delta &=& \delta_N \, ,
\end{eqnarray}
together with $B^i = h_{ij} = 0$. 
This simplified dictionary is very commonly assumed.
However, on scales comparable
to the Hubble radius, all terms in \eqref{eq:dict-L3} should be of 
comparable size, so if one is interested in investigating
behavior on large scales, the full dictionary or abridged dictionary should
be used. 

As explained above, our dictionary
\eqref{eq:dict-L1}--\eqref{eq:dict-L5} has been constructed so as to
produce a solution of Einstein's equation to linearized order in $(A,
B_i, H_L, h_{ij}, \delta, v^i)$. A Newtonian cosmology that
corresponds to our universe should have $\psi_N \ll 1$ and $|v_N^i|
\ll 1$, but will normally have $\delta_N \gg 1$ on small
scales. Therefore, it is not obvious, {\it a priori} how accurate our
dictionary is in producing a solution to Einstein's equation.  In
fact, it is clear that there may be difficulties in this regard
because the dictionary should produce a spacetime that nearly
satisfies the linearized Einstein equation on small scales, but the
linearized Einstein equation is incompatible (via the linearized
Bianchi identity) with the nonlinear dynamical behavior of
matter\footnote{In many references (see, e.g.,
  \cite{Bertschinger:1993xt,Chisari:2011iq}), the linearized Einstein
  equation is written down together with the nonlinear dynamical
  equations for matter. This combined system of equations is
  mathematically inconsistent.}  that occurs on small scales. We now
investigate how close \eqref{eq:dict-L1}--\eqref{eq:dict-L5} comes to
producing a solution to Einstein's equation.

\section{Counting scheme and improved dictionary}\label{sec:improveddictionary}

As stated at the end of the previous section, we wish to determine how
close our dictionary comes to producing a solution of Einstein's
equation when the Newtonian cosmology has $\psi_N \ll 1$ and $|v_N^i|
\ll 1$, but may have $\delta_N \gg 1$ at small scales. To analyze this
issue, we need a consistent approximation scheme that can take
advantage of the fact that the deviation of the metric from an FLRW
model is small on all scales, but permits very large deviations of the
stress-energy tensor from an FLRW model to occur on small scales.
Such an approximation scheme was recently developed by us in
\cite{Green:2010qy} and used to analyze the backreaction effects of
small scale inhomogeneities on large scale dynamics. We refer the
reader to that reference for the precise mathematical formulation of
the approximation. For our purposes here, it suffices to observe that,
as in ordinary perturbation theory, in our approximation scheme there
is a ``small parameter'' $\epsilon$ (denoted $\lambda$ in
\cite{Green:2010qy}) that measures the deviation, $\gamma_{ab} =
g_{ab} - g^{(0)}_{ab}$, of the metric $g_{ab}$ from a background
metric $g^{(0)}_{ab}$, so $\gamma_{ab} = O(\epsilon)$. However, unlike
ordinary perturbation theory, first spacetime derivatives of
$\gamma_{ab}$ are allowed to be $O(1)$, and second spacetime
derivatives of $\gamma_{ab}$---and, hence, the deviations of the
stress-energy tensor from the background stress-energy---are allowed
to be $O(1/\epsilon)$.  In particular, the quadratic products
$\nabla_c \gamma_{ab} \nabla_f \gamma_{de}$ and $\gamma_{ab}
\nabla_c\nabla_f \gamma_{de}$ that appear in Einstein's equation are
$O(1)$, so our approximation scheme allows small scale inhomogeneities
to affect the dynamics of the background metric. One of the main
results of \cite{Green:2010qy} is that, in fact, the only possible
effect that these nonlinear terms can have on the dynamics of the
background metric is to contribute an effective stress-energy that is
traceless and has positive energy, corresponding to the presence of
gravitational radiation.  For the present work, we assume that the
universe contains a negligible amount of gravitational radiation, so
that this effective stress-energy tensor can be set to zero, and the
background metric (which has FLRW symmetry) therefore obeys the
ordinary Einstein equation with dust stress energy tensor.

In addition to analyzing the effects of small scale inhomogeneities on
the dynamics of the background metric, in \cite{Green:2010qy}
perturbation theory was generalized to allow for significant
nonlinearity at small scales, while at the same time maintaining a
linearized description at large scales (see also
\cite{Baumann:2010tm}).  In order to ascribe different behavior to
perturbations at different scales these notions must of course be
defined.  In \cite{Green:2010qy}, the notion of the ``long wavelength
part'' of quantities was defined in a mathematically precise manner by
considering the weak limit of these quantities as $\epsilon
\rightarrow 0$. As explained in \cite{Green:2010qy}, at sufficiently
small but finite $\epsilon$, this should correspond closely to taking
an average over a spatial scale\footnote{For the present universe,
  $L\approx100\text{ Mpc}$ should meet these criteria.} $L$ that is
small compared with the background curvature (i.e., the Hubble radius)
but sufficiently large that at this scale and beyond we have
$|\delta|\ll1$. For the present work, we shall identify the long
wavelength part, $A_{a_1\cdots a_n}^{(L)} $, of a tensor field,
$A_{a_1\cdots a_n}$, with the spatial average\footnote{In
  \cite{Green:2010qy} averages over (short) time intervals were also
  performed. The purely spatial averaging we perform here with a
  suitable ``window function'' corresponds precisely to the averaging
  done in \cite{Baumann:2010tm}.} of its components
\begin{equation}
 A_{\mu_1\cdots \mu_n}^{(L)} (x) =  \langle A_{\mu_1\cdots \mu_n}\rangle(x) \equiv 
 \int d^3x'\,W_L(x-x')A_{\mu_1\cdots \mu_n}(x')\,,
\label{windfun}
\end{equation}
using a suitable ``window function,'' $W_L(x-x')$, of size $L$, i.e.,
a smooth function which is equal to 1 for $a^2|x-x'|^2<L$, and which
smoothly falls to 0 outside of this region\footnote{Equivalently, one
  could work in Fourier transform space and multiply the Fourier
  transform, $ \hat{A}_{a_1\cdots a_n}^{(L)} (k)$, by $\hat{W}_L(k)$,
  where $\hat{W}_L(k)$ interpolates between $1$ (for $k< 1/L$) and $0$
  for $k \gg 1/L$.}.  The requirement that $L$ be much smaller than
the background curvature scale ensures that this averaging process is
well-defined, whereas the requirement that $L$ be sufficiently large
that $|\delta|\ll1$ should ensure that the long wavelength parts of
perturbations behave linearly.

We define the ``short wavelength part'' of $A_{a_1\cdots a_n}$ by
\begin{equation}
 A_{a_1\cdots a_n}^{(S)} \equiv A_{a_1\cdots a_n} - A_{a_1\cdots a_n}^{(L)} \, ,
\end{equation}
thereby providing a decomposition of any quantity into its long and short
wavelength parts.
The framework of \cite{Green:2010qy}
allows one to make different assumptions in a mathematically
consistent manner about the long and short
wavelength parts of the various quantities. In particular, 
derivatives of short wavelength parts can pick up the
factors of $1/\epsilon$ described above, but derivatives do not increase the size
of long wavelength parts.

Our framework can be straightforwardly applied to cosmological
Newtonian gravity. It is natural in this case also to impose the
additional conditions that velocities are suitably ``small'' and time
derivatives of quantities are correspondingly small compared with
space derivatives at small scales.  Specifically, the sizes we
assign\footnote{The orders we assign to the quantities in Table
  \ref{tab:Norders} correspond to the post-Newtonian orderings of
  Futamase and Schutz \cite{Futamase:1983} up to a rescaling of the
  spatial coordinates, but they differ from the post-Newtonian
  orderings of Oliynyk \cite{Oliynyk:2009bc,Oliynyk:2009qa}.}  to the
short wavelength part of the quantities $(\psi_N, v^i_N, \delta_N)$ of
the previous section are given in Table \ref{tab:Norders}.
\begin{table}[htb]
  \centering
  \begin{tabular}{c|c}
    Quantity & Order \\
    \hline
    $\psi_N^{(S)}$ & $\epsilon$ \\
    $v_N^{(S) i}$ & $\epsilon^{1/2}$ \\
    $\delta_N^{(S)}$ & $\epsilon^{-1}$ \\
    $\partial_0$ & $\epsilon^{-1/2}$ \\
    $\partial_i$ & $\epsilon^{-1}$
  \end{tabular}
  \caption{Small scale order counting for Newtonian quantities}
  \label{tab:Norders}
\end{table}
On the other hand, the long wavelength part
of all of these quantities and their space and time derivatives are assumed to be
$O(\epsilon)$. It should be noted that 
certain products of short wavelength quantities can have
$O(\epsilon)$ large scale average.  In particular, long wavelength averages of nonlinear
quantities corresponding Newtonian potential energy\footnote{\label{lemma}A
  priori, $\langle\delta_N\psi_N\rangle=O(1)$, but due to the fact
  that $\delta_N$ is bounded below by $-1$, in fact
  $\langle\delta_N\psi_N\rangle=O(\epsilon)$; see the lemma of section
  II of \cite{Green:2010qy}.}, kinetic energy, and
linear momentum enter the perturbation equation for $\gamma_{ab}^{(L)}$.

Our aim here is simply to use the framework of \cite{Green:2010qy} as
a ``counting scheme'' in powers of $\epsilon$ to see how close our
dictionary comes to producing a solution to Einstein's equation.
Specifically, we assume that we have been provided with a Newtonian
cosmological solution where the ``sizes'' of quantities correspond to
Table \ref{tab:Norders}. To complete our ``counting scheme,'' we must
also assign an $\epsilon$-order to $B_i^{(S)}$. Since $B_i$ is
obtained by solving the Poisson equation \eqref{eq:dict-L5} and,
according to Table \ref{tab:Norders}, the source term is of order
$\epsilon^{-1/2}$, we assign $B_i^{(S)}$ the order $\epsilon^{3/2}$.

Having assigned $\epsilon$-orders to all quantities,
we may ask the following question: If we substitute
the Newtonian solution into our 
dictionary \eqref{eq:dict-L1}--\eqref{eq:dict-L5}
to produce a spacetime metric $g_{ab}$ and dust stress-energy tensor
$T_{ab}$, how close does $(g_{ab}, T_{ab})$ come to satisfying 
Einstein's equation?

If we are to have confidence that the dictionary is producing a good
approximation to a solution to Einstein's equation, we would want
Einstein's equation to be solved to at least $O(1)$ in
$\epsilon$. This is a nontrivial requirement, since there are
individual terms, such as $\delta$, in Einstein's equation that are
$O(1/\epsilon)$ in our counting scheme. As we shall see below, the
dictionary \eqref{eq:dict-L1}--\eqref{eq:dict-L5} solves Einstein's
equation to $O(1/\epsilon)$ but fails to yield a solution to
Einstein's equation at $O(1)$ in $\epsilon$. Nevertheless, we will
then show that we can make further small corrections to the metric so
that Einstein's equation does hold to $O(1)$. As we shall see, these
metric corrections should be $O(\epsilon^2)$ at small scales and
therefore should be negligible. If so, our original dictionary
\eqref{eq:dict-L1}--\eqref{eq:dict-L5} should be producing an accurate
relativistic cosmology in terms of its description of the metric and
matter distribution on small scales.

In addition, if the dictionary is to be trusted for its description of
large scale structure---including on scales comparable to (or larger
than) the Hubble radius---we would want Einstein's equation to hold to
at least $O(\epsilon)$ at large scales. As we shall see below, even
after we have made the necessary corrections to the metric so that
Einstein's equation is satisfied to $O(1)$ at small scales, Einstein's
equation will fail to hold to $O(\epsilon)$ at large scales in our
counting scheme. We will therefore make further large scale
corrections to the dictionary so that Einstein's equation holds to
$O(\epsilon)$ at large scales. As we shall see, although these
corrections are formally of order $\epsilon$, they would be expected
to make negligible corrections to ordinary linearized perturbation
theory at long wavelengths. If so, our original dictionary
\eqref{eq:dict-L1}--\eqref{eq:dict-L5} should be producing an accurate
relativistic cosmology in terms of its description of the metric and
matter distribution on large scales.

The above corrections provide us with an improved dictionary that
incorporates the dominant general relativistic corrections to
\eqref{eq:dict-L1}--\eqref{eq:dict-L5}.  Although the improved
dictionary is undoubtedly far more precise than would be needed for
most applications, it is important as a matter of principle to know
that corrections can be made so that Einstein's equation holds to
$O(1)$ at all scales and at $O(\epsilon)$ on large
scales. Furthermore, for any given Newtonian cosmology, the correction
terms appearing in the improved dictionary can be calculated
straightforwardly, and their size should give a reliable indication of
the accuracy of the original dictionary
\eqref{eq:dict-L1}--\eqref{eq:dict-L5}. If, as indicated above, these
correction terms are negligibly small, then the Newtonian cosmology
should provide---via the original dictionary
\eqref{eq:dict-L1}--\eqref{eq:dict-L5} and/or its abridgment
\eqref{eq:adict-L1}--\eqref{eq:adict-L3} or simplification
\eqref{eq:sdict-L1}--\eqref{eq:sdict-L3}---an excellent description of
what is predicted by general relativity.

\subsection{Solving Einstein's equation to $O(1)$}

\subsubsection{How well are Einstein's equation solved by the original dictionary?}

Appendix \ref{sec:Einstein} presents the calculation of Einstein's
equation for the metric \eqref{eqn:Poissongauge} and stress-energy
tensor \eqref{Tab}--\eqref{delta}, keeping all terms that could
potentially contribute to $O(1)$ as well as all terms that could
potentially contribute to $O(\epsilon)$ at large scales.  Given a
Newtonian cosmological solution $(\psi_N, v^i_N, \delta_N)$, we
substitute it into the dictionary
\eqref{eq:dict-L1}--\eqref{eq:dict-L5}, and substitute the result into
Einstein's equation, freely using the Newtonian equations to simplify
the resulting expressions.  Equation~\eqref{eqn:E00} yields
\begin{eqnarray}
  &&G^0_{\phantom{0}0}(g) + \Lambda - 8\pi T^0_{\phantom{0}0}\nonumber\\
  &=&\frac{3}{a^2}\left\{-\frac{2}{3}\partial^i\partial_i\psi_N-\frac{8}{3}\psi_N\partial^i\partial_i\psi_N-\partial^i\psi_N\partial_i\psi_N\right\}-8\pi\rho_0\left\{-\delta_N-(1+\delta_N)v_N^iv_{Ni}\right\}+o(1)\nonumber\\
  &=&\frac{3}{a^2}\left\{-\frac{8}{3}\psi_N\partial^i\partial_i\psi_N-\partial^i\psi_N\partial_i\psi_N\right\}+8\pi\rho_0(1+\delta_N)v_N^iv_{Ni}+o(1)\, ,
\end{eqnarray}
where we used the Poisson equation
\eqref{eq:NG-pert} for $\psi_N$ in the second equality. 
Since the quantities $\psi_N\partial^i\partial_i\psi_N$, $\partial^i\psi_N\partial_i\psi_N$
and $\delta_N v_N^iv_{Ni}$ are each
$O(1)$ in our counting scheme (and these terms do not cancel), we see that this component
of Einstein's equation is {\em not} satisfied to $O(1)$.

Equation~\eqref{eqn:E0i} yields
\begin{equation}
  G^0_{\phantom{0}i}(g) - 8\pi T^0_{\phantom{0}i}
  =\frac{2}{a^2}\left\{-\partial_i\dot{\psi}_N-\frac{\dot{a}}{a}\partial_i\psi_N\right\}-8\pi\rho_0\left.\left[(1+\delta_N)v_{Ni}\right]\right|_{\text{s}}+o(1)\,.
\end{equation}
However, from
the Newtonian equations of motion \eqref{eq:NG-pert} and
\eqref{eq:Nmass-pert}, it follows that
\begin{equation}\label{eq:LdotpsiN}
  \partial^i\partial_i\dot{\psi}_N+\frac{\dot{a}}{a}\partial^i\partial_i\psi_N=-4\pi \rho_0a^2 \partial_i\left[(1+\delta_N)v_N^i\right]\,.
\end{equation}
On the torus this may be integrated, giving
\begin{equation}\label{eq:LdotpsiN2}
  \partial_i\dot{\psi}_N+\frac{\dot{a}}{a}\partial_i\psi_N=-4\pi\rho_0a^2\left.\left[(1+\delta_N)v_{Ni}\right]\right|_{\text{s}}\,.
\end{equation}
Thus, we obtain
\begin{equation}
  G^0_{\phantom{0}i}(g) - 8\pi T^0_{\phantom{0}i} =  o(1)\,,
\end{equation}
i.e., these components of Einstein's equation {\em are} satisfied\footnote{In fact, the precise forms of
\eqref{eq:dict-L2} and \eqref{eq:dict-L5} were chosen so that no further corrections to
$v^i$ and $B^i$ would be needed to satisfy Einstein's equation to $O(1)$.} to $O(1)$.

Finally, from eq.~\eqref{eqn:Eij} we obtain the space-space components, 
\begin{eqnarray}
  &&G^i_{\phantom{i}j}(g) + \Lambda\delta^i_{\phantom{i}j}- 8\pi T^i_{\phantom{i}j}\nonumber\\
  &=&\frac{1}{a^2}\left\{2\ddot{\psi}_N-4\psi_N\partial^k\partial_k\psi_N-3\partial_k\psi_N\partial^k\psi_N\right\}\delta^i_{\phantom{i}j}+\frac{1}{a^2}\left\{4\psi_N\partial^i\partial_j\psi_N+2\partial^i\psi_N\partial_j\psi_N\right\}\nonumber\\
  &&+\frac{1}{2a^2}\left\{\partial^i\dot{B}_j+\partial_j\dot{B}^i\right\}-8\pi\rho_0(1+\delta_N)v_N^iv_{Nj}+o(1)\,.
\end{eqnarray}
Thus, these components of Einstein's equation are {\em not} satisfied to $O(1)$.

\subsubsection{Corrections to the dictionary needed to solve Einstein's equation to $O(1)$}\label{sec:O1}

We will now show that all components of Einstein's equation can be
satisfied to $O(1)$ by making the additional corrections $\chi$,
$\xi$, and $\jmath_{ij}$ to the spacetime metric as follows:
\begin{eqnarray}
  \label{eqn:d3}A&=&\psi_N+\chi+\xi\,,\\
  \label{eqn:d4}H_L&=&-\psi_N-\chi\,,\\
  \label{eqn:d6}
  h_{ij} &=& \jmath_{ij}\,,
\end{eqnarray}
with $\xi^{(S)}$, $\chi^{(S)}$, and $\jmath_{ij}^{(S)}$ all $O(\epsilon^2)$.
However, we do not make any modifications to the original dictionary expressions for
$v^i$, $\delta$, and $B^i$, i.e., we continue to use
\begin{eqnarray}
 \label{eqn:d7} (1+\delta_N)v^i&=&(1+\delta_N)(v^i_N+ B^i)-\left.\overline{(1+\delta_N)v_N^i}\right|_{\text{v}}\,,\\
\label{eqn:d8}\delta &=& \delta_N-\frac{3}{4\pi\rho_0a^2}\left[\left(\frac{\dot{a}}{a}\right)^2\psi_N+\frac{\dot{a}}{a}\dot{\psi}_N\right]\,,\\
\label{eqn:d9}\partial^j\partial_j B^i &=& -16\pi\rho_0a^2\left.\left((1+\delta_N)v_N^i-\overline{(1+\delta_N)v_N^i}\right)\right|_{\text{v}}\, .
\end{eqnarray}
In particular, it should be emphasized that no additional corrections are made to the
matter distribution variables $\delta$ and $v^i$.

We have already seen that the original dictionary
solved the time-space components of Einstein's equation to $O(1)$ 
and it is not difficult to see that these
equations continue to hold with the above revisions. Thus, to solve
Einstein's equation to $O(1)$, we need only consider
the space-space components \eqref{eqn:Eij} and the time-time component  \eqref{eqn:E00}.
To solve \eqref{eqn:Eij}, we note that we can 
uniquely decompose any 
symmetric tensor field $E_{ij}$ on a $3$-torus 
with flat metric $\delta_{ij}$ and flat derivative operator $\partial_i$ as
\begin{equation}
E_{ij} = U \delta_{ij} + 
\partial_i \partial_j V - \frac{1}{3}\partial^k \partial_k V
+ 2\partial_{(i} W_{j)} +X_{ij} \, ,
\label{svt}
\end{equation}
with $\partial^i W_i = 0$, $\partial^i X_{ij} = 0$, and ${X^i}_i = 0$. This
defines the decomposition of $E_{ij}$ into its scalar ($U,V$), vector ($W_i$),
and tensor ($X_{ij}$) parts. Thus, we can solve an equation of the form
$E_{ij} = 0$ by separately solving its scalar, vector, and tensor parts.
To begin, we take the 
double divergence of the traceless part of \eqref{eqn:Eij}. 
We obtain\footnote{In
  substituting for $v^i$ we have neglected some terms proportional to
  $1/(1+\delta_N)$, which can in fact be quite large in low density
  regions.  When one makes a uniform momentum correction
  $-\left.\overline{(1+\delta_N)v_N^i}\right|_{\text{v}}$, this
  corresponds, in a low density region, to a very large velocity
  correction which is unphysical.  If such a situation were to occur,
  then a fix would be to transfer some of this momentum to a higher
  density region.}
\begin{eqnarray}\label{eqn:xi}
  &&-\frac{2}{3a^2}\partial^i\partial_i\partial^j\partial_j\xi+\frac{1}{a^2}\partial_i\partial^j\left\{4\psi_N\partial^i\partial_j\psi_N+2\partial^i\psi_N\partial_j\psi_N\right\}\nonumber\\
  &&-\frac{1}{3a^2}\partial^i\partial_i\left\{4\psi_N\partial^j\partial_j\psi_N+2\partial^j\psi_N\partial_j\psi_N\right\}\nonumber\\
  &=&8\pi\rho_0\partial_i\partial^j\left[(1+\delta_N)v_N^iv_{Nj}\right]-\frac{8\pi}{3}\rho_0\partial^i\partial_i\left[(1+\delta_N)v_N^kv_{Nk}\right]+o\left(\frac{1}{\epsilon^2}\right)\,.
\end{eqnarray}
Here we have dropped terms which are $o(1/\epsilon^2)$, since we have taken two
spatial derivatives of an equation that we wish to satisfy to $O(1)$.
We can solve \eqref{eqn:xi} to the desired order by defining $\xi$ to
be the solution to the following double Poisson equation:
\begin{eqnarray}\label{eq:xidef}
  \partial^i\partial_i\partial^j\partial_j\xi&=&3\partial_i\partial^j\left\{2\psi_N\partial^i\partial_j\psi_N+\partial^i\psi_N\partial_j\psi_N\right\}-\partial^i\partial_i\left\{2\psi_N\partial^j\partial_j\psi_N+\partial^j\psi_N\partial_j\psi_N\right\}\nonumber\\
  &&-12\pi\rho_0a^2\partial_i\partial^j\left[(1+\delta_N)v_N^iv_{Nj}\right]+4\pi\rho_0a^2\partial^i\partial_i\left[(1+\delta_N)v_N^kv_{Nk}\right]\,.
\end{eqnarray}
A solution for $\xi$ exists on a torus because the source term is a
divergence and therefore has no spatially constant piece.  This
solution is unique up to a spatially constant function of time, which
we fix by requiring that its spatial average, $\overline{\xi}$,
vanishes.  Since the double divergence of the traceless part of
\eqref{eqn:Eij} has now been solved to $O(1/\epsilon^2)$, the scalar
part of the traceless part of \eqref{eqn:Eij} should now be solved to
$O(1)$, as desired. Note that since the four spatial derivatives
applied to $\xi$ yields a quantity that is $O(1/\epsilon^2)$, the
short wavelength part, $\xi^{(S)}$, of $\xi$ should be
$O(\epsilon^2)$, so our assumption that $\xi^{(S)}$ is $O(\epsilon^2)$
is self-consistent.

Next, we show that, with this choice of $\xi$, the trace of
\eqref{eqn:Eij} also is satisfied to $O(1)$.  Substituting the revised
dictionary \eqref{eqn:d3}--\eqref{eqn:d9} into the trace of
\eqref{eqn:Eij}, we find that we must satisfy
\begin{eqnarray}\label{eq:Eii1}
  &&\frac{3}{a^2}\left\{\frac{2}{3}\partial^i\partial_i\xi+2\ddot{\psi}_N\right\}+\frac{1}{a^2}\left\{-8\psi_N\partial^i\partial_i\psi_N-7\partial_i\psi_N\partial^i\psi_N\right\}\nonumber\\
  &=&8\pi\rho_0(1+\delta_N)v_N^iv_{Ni}+o(1)\,.
\end{eqnarray}
To see if this equation holds, we take its Laplacian. The double
Laplacian of $\xi$ will then appear, and we can substitute for this
quantity using \eqref{eq:xidef}.  Since we want to solve
\eqref{eq:Eii1} to $O(1)$, and each spatial derivative increases the
small scale order by a factor of $1/\epsilon$, we wish to solve the
Laplacian of \eqref{eq:Eii1} to $O(1/\epsilon^2)$, so the equation we
wish to solve is
\begin{equation}
  \frac{3}{a^2}\left\{2\partial^i\partial_i\ddot{\psi}_N-2\partial^i\left(\partial_i\psi_N\partial^j\partial_j\psi_N\right)\right\}=24\pi\rho_0\partial^j\partial_i\left[(1+\delta_N)v_N^iv_{Nj}\right]+o\left(\frac{1}{\epsilon^2}\right)\,.
\label{tr}
\end{equation}
However, using the Newtonian equations
\eqref{eq:NG-pert}--\eqref{eq:NEuler-pert}, as well as the Friedmann
equations for the Newtonian background, one can show that
\begin{eqnarray}\label{eq:ddotpsi-identity}
  &&\partial^i\partial_i\left\{2\ddot{\psi}_N+6\frac{\dot{a}}{a}\dot{\psi}_N+\left[4\partial_\tau\left(\frac{\dot{a}}{a}\right)+2\left(\frac{\dot{a}}{a}\right)^2\right]\psi_N\right\}\nonumber\\
  &=&8\pi\rho_0a^2\partial_i\partial_j\left[(1+\delta_N)v_N^iv_N^j\right]+2\partial_i\left(\partial^j\partial_j\psi_N\partial^i\psi_N\right)\,,
\end{eqnarray}
so \eqref{tr} is indeed solved to
the desired order.  We have thus fully solved the scalar parts of the space-space part of
Einstein's equation to $O(1)$.

Next, we consider the vector part of \eqref{eqn:Eij} by taking its
divergence, using the fact that the scalar parts have already been
solved.  We obtain
\begin{eqnarray}\label{eq:BO1}
  &&\frac{1}{2a^2}\partial^j\partial_j\dot{B}^i+\frac{1}{a^2}\partial^j\left.\left\{4\psi_N\partial^i\partial_j\psi_N+2\partial^i\psi_N\partial_j\psi_N\right\}\right|_{\text{v}}\nonumber\\
  &=&\left.8\pi\rho_0\partial^j\left[(1+\delta_N)v_N^iv_{Nj}\right]\right|_{\text{v}}+o\left(\frac{1}{\epsilon}\right)\,.
\end{eqnarray}
Using the Newtonian Euler and mass conservation equations,
along with the definition of $B_i$, one can show that
\begin{eqnarray}\label{eq:BN-identity}
  &&\frac{1}{2a^2}\left\{\partial^j\partial_j\dot{B}^i+2\frac{\dot{a}}{a}\partial^j\partial_jB^i\right\}+\frac{1}{a^2}\partial^j\left.\left\{4\psi_N\partial^i\partial_j\psi_N+2\partial^i\psi_N\partial_j\psi_N\right\}\right|_{\text{v}}\nonumber\\
  &=&\left.8\pi\rho_0\partial^j\left[(1+\delta_N)v_N^iv_{Nj}\right]\right|_{\text{v}}\,,
\end{eqnarray}
so equality does hold for the terms explicitly written in
\eqref{eq:BO1}.  Thus, the vector part of \eqref{eqn:Eij} is satisfied
to $O(1)$.

The tensor part of \eqref{eqn:Eij} is all that remains of this equation. 
We obtain
\begin{equation}
  -\frac{1}{a^2}\partial^k\partial_k\jmath^i_{\phantom{i}j}+\frac{1}{a^2}\left.\left\{4\psi_N\partial^i\partial_j\psi_N+2\partial^i\psi_N\partial_j\psi_N\right\}\right|_{\text{t}}
  =\left.8\pi\rho_0(1+\delta_N)v_N^iv_{Nj}\right|_{\text{t}}+o(1)\, ,
\label{tp}
\end{equation}
where $|_t$ denotes the tensor part of a quantity in its decomposition 
\eqref{svt}. To solve this equation 
to the desired order, we define $\jmath_{ij}$ to be the solution of
\begin{eqnarray}\label{eq:h1}
  \partial^k\partial_k\jmath^i_{\phantom{i}j}&=&\left.\left\{8\pi\rho_0a^2(1+\delta_N)v_N^iv_{Nj}+4\psi_N\partial^i\partial_j\psi_N+2\partial^i\psi_N\partial_j\psi_N\right.\right.\nonumber\\
  &&\left.\left.-8\pi\rho_0a^2\overline{(1+\delta_N)v_N^iv_{Nj}}+2\overline{\partial^i\psi_N\partial_j\psi_N}\right\}\right|_{\text{t}}\,,
\end{eqnarray}
where the overline denotes spatial average (see \eqref{bar}). The terms with the overline
in \eqref{eq:h1} have been added in so that the source has vanishing integral, as is necessary 
in order to be able to solve the Poisson equation. Since these terms 
are\footnote{See footnote \ref{lemma}.} $O(\epsilon)$, we will satisfy \eqref{tp}
to the desired order by choosing $\jmath_{ij}$
to solve \eqref{eq:h1}. We fix the ambiguity in $\jmath_{ij}$ by requiring 
$\overline{\jmath}_{ij}=0$.  Since two spatial derivatives applied to $\jmath_{ij}$
yields a quantity that is $O(1)$, the short wavelength part, $\jmath_{ij}^{(S)}$, of $\jmath_{ij}$
should be $O(\epsilon^2)$, so our assumption that $\jmath_{ij}^{(S)}$, is $O(\epsilon^2)$ is self-consistent.
We have now solved \eqref{eqn:Eij} to $O(1)$.

Finally, we consider the time-time component of Einstein's equation.
Substitution of the dictionary into \eqref{eqn:E00} yields
\begin{eqnarray}
  &&\frac{3}{a^2}\left\{-\frac{2}{3}\partial^i\partial_i\psi_N-\frac{2}{3}\partial^i\partial_i\chi-\frac{8}{3}\psi_N\partial^i\partial_i\psi_N-\partial^i\psi_N\partial_i\psi_N\right\}\nonumber\\
  &=&8\pi\rho_0\left[-\delta_N-(1+\delta_N)v_N^iv_{Ni}\right]+o(1)\,.
\end{eqnarray}
Using the Newtonian field equation \eqref{eq:NG-pert}, we obtain
\begin{equation}\label{eq:E002}
  \frac{3}{a^2}\left\{-\frac{2}{3}\partial^i\partial_i\chi-\frac{8}{3}\psi_N\partial^i\partial_i\psi_N-\partial^i\psi_N\partial_i\psi_N\right\}
  =-8\pi\rho_0(1+\delta_N)v_N^iv_{Ni}+o(1)\,.
\end{equation}
We define $\chi$ to be the solution to the Poisson equation
\begin{eqnarray}\label{eq:chi1}
  \partial^i\partial_i\chi&=&-4\psi_N\partial^i\partial_i\psi_N-\frac{3}{2}\partial^i\psi_N\partial_i\psi_N+4\pi a^2\rho_0(1+\delta_N)v_N^iv_{Ni}\nonumber\\
  &&-\frac{5}{2}\overline{\partial^i\psi_N\partial_i\psi_N}-4\pi a^2\rho_0\overline{(1+\delta_N)v_N^iv_{Ni}}\,,
\end{eqnarray}
with $\overline{\chi}=0$.  Since two spatial derivatives applied to
$\chi$ yields a quantity that is $O(1)$, the short wavelength part,
$\chi^{(S)}$, of $\chi$ should be $O(\epsilon^2)$.

Thus, we have shown that Einstein's equation can be solved to $O(1)$
by making the corrections \eqref{eqn:d3}--\eqref{eqn:d6} to the
original dictionary, where $\chi$, $\xi$, and $\jmath_{ij}$ are given,
respectively, by \eqref{eq:chi1}, \eqref{eqn:xi}, and
\eqref{eq:h1}. Although it is extremely important as a matter of
principle that such corrections can be made so as to obtain a solution
to $O(1)$, we expect that these corrections will be negligibly small
compared with $\psi_N$.

\subsection{Improving the solution to $O(\epsilon)$ at large scales}

In the previous subsection we obtained a solution to $O(1)$. However,
as previously stated, 
if our dictionary is to be trusted for its description of
large scale structure---including on scales comparable to (or larger
than) the Hubble radius---we want Einstein's equation to hold to
at least $O(\epsilon)$ at large scales. 
Within the context of ordinary
perturbation theory, this corresponds to solving the linearized
perturbation equation.  Within our generalized perturbative
framework, this corresponds to solving the generalized linearized
perturbation equation (87) of \cite{Green:2010qy}.  The difference
between these, as
noted earlier, is that long wavelength averages of products of small
scale quantities enter into the generalized linearized
equation.

It is easy to check that, even with the corrections 
\eqref{eqn:d3}--\eqref{eqn:d6}, our dictionary does {\em not} produce a solution
to $O(\epsilon)$ at long wavelengths. 
Therefore, we will need to make the following additional
{\em long wavelength} corrections to our metric and
matter variables:
\begin{eqnarray}
  \label{eqn:d1A}A&=&\psi_N+\chi+\xi+X+\Xi\,,\\
  \label{eqn:d2A}H_L&=&-\psi_N-\chi-X\,,\\
  \label{eqn:d3A}(1+\delta_N)v_i&=&(1+\delta_N)(v_{Ni}+ B_i)-\left.\overline{(1+\delta_N)v_{Ni}}\right|_{\text{v}}+P_i\,,\\
  \label{eqn:d4A}\delta&=&\delta_N-\frac{3}{4\pi\rho_0a^2}\left[\left(\frac{\dot{a}}{a}\right)^2\psi_N+\frac{\dot{a}}{a}\dot{\psi}_N\right]+\Delta\,,\\
  \label{eqn:d5A}h_{ij}&=&\jmath_{ij}+J_{ij}\,.
\end{eqnarray}
No additional long wavelength correction is needed for $B_i$. Here, the quantities
$\Xi$, $X$, $P_i$, $\Delta$, and $J_{ij}$ are assumed to be $O(\epsilon)$
and to have vanishing short wavelength part. Hence, they do not contribute
to Einstein's equation to $O(1)$ and thus do not spoil the solution
obtained in the previous subsection.

Our strategy is to apply the averaging operator
$\langle\cdot\rangle$ (see eq.~\eqref{windfun}) 
to Einstein's equation, and to choose the above new
correction terms in order to obtain a solution to $O(\epsilon)$. 
For our the calculations below, it is useful to 
note that the averaging operator $\langle\cdot\rangle$ commutes
with differentiation. Note also that since $\psi_N = O(\epsilon)$,
we clearly have $\langle \psi^2_N \rangle = O(\epsilon^2)$,
and, consequently, we have
\begin{equation}
\langle \partial^i \psi_N \partial_i \psi_N \rangle + 
\langle \psi_N \partial^i \partial_i \psi_N \rangle = \frac{1}{2}
\partial^i \partial_i \langle \psi^2_N \rangle = O(\epsilon^2) \, .
\end{equation}
Thus, we may freely ``integrate by parts'' to set 
$\langle \psi_N \partial^i \partial_i \psi_N \rangle =
- \langle \partial^i \psi_N \partial_i \psi_N \rangle$ in our calculations.

As before, we begin with the double divergence of the trace-free part of
the space-space components \eqref{eqn:Eij} of
Einstein's equation. Substituting our new dictionary 
\eqref{eqn:d1A}--\eqref{eqn:d5A}, applying
the averaging operator $\langle\cdot\rangle$, 
and using the equation obtained by applying the averaging operator to \eqref{eq:xidef} to simplify 
the resulting expression, we obtain
\begin{equation}\label{eq:Xidef}
  -\frac{2}{3a^2}\partial^i\partial_i\partial^j\partial_j\Xi=o(\epsilon)\,.
\end{equation}
Thus $\Xi$ can
only have a spatially constant part, i.e.,
\begin{equation}
  \Xi = \overline{\Xi}\,,
\label{Xiconst}
\end{equation}
where $\overline{\Xi}$ may be an arbitrary function of $\tau$. Examining the
scalar homogeneous parts of the metric,
\begin{equation}\label{eq:metricconst}
  \overline{ds}^2=a^2(\tau)\left[-(1+2\overline{X}+2\overline{\Xi})d\tau^2 + (1-2\overline{X})\delta_{ij}dx^idx^j\right]\,,
\end{equation}
we see that $\overline{\Xi}$ corresponds to gauge freedom in the choice 
of time coordinate. We fix this freedom by setting
\begin{equation}\label{eq:Xidef2}
  \Xi=-2\overline{X}\,,
\end{equation}
corresponding to using conformal time.

Next, we consider the trace of \eqref{eqn:Eij}.
Substituting from the dictionary, applying $\langle\cdot\rangle$, 
and using $\Xi=-2\overline{X}$, we obtain
\begin{eqnarray}\label{eq:Eiismooth}
  &&\frac{3}{a^2}\left\{\frac{2}{3}\partial^i\partial_i\langle\xi\rangle + 2\left(\langle\ddot{\psi}_N\rangle+\langle\ddot{\chi}\rangle+\ddot{X}\right) + 2\frac{\dot{a}}{a}\left(3\langle\dot{\psi}_N\rangle+3\langle\dot{\chi}\rangle+3\dot{X}+\langle\dot{\xi}\rangle-2\dot{\bar{X}}\right)\right.\nonumber\\
  &&\left.+\left[4\partial_\tau\left(\frac{\dot{a}}{a}\right)+2\left(\frac{\dot{a}}{a}\right)^2\right]\left(\langle\psi_N\rangle+\langle\chi\rangle+X+\langle\xi\rangle -2\bar{X}\right)\right\} + \frac{1}{a^2}\langle\partial^i\psi_N\partial_i\psi_N\rangle\nonumber\\
  &=&8\pi\rho_0\langle\left(1+\delta_N\right)v_N^iv_{Ni}\rangle+o(\epsilon)\,.
\end{eqnarray}
As before, we take the Laplacian of this equation and
substitute the average of \eqref{eq:xidef}, obtaining
\begin{eqnarray}
  &&\frac{3}{a^2}\partial^j\partial_j\left\{2\left(\langle\ddot{\psi}_N\rangle+\langle\ddot{\chi}\rangle+\ddot{X}\right) + 2\frac{\dot{a}}{a}\left(3\langle\dot{\psi}_N\rangle+3\langle\dot{\chi}\rangle+3\dot{X}+\langle\dot{\xi}\rangle \right)\right.\nonumber\\
  &&\left.+\left[4\partial_\tau\left(\frac{\dot{a}}{a}\right)+2\left(\frac{\dot{a}}{a}\right)^2\right]\left(\langle\psi_N\rangle+\langle\chi\rangle+X+\langle\xi\rangle\right)\right\} - \frac{6}{a^2}\partial_i\langle\partial^i\psi_N\partial^j\partial_j\psi_N\rangle\nonumber\\
  &=&24\pi\rho_0\partial_i\partial^j\langle\left(1+\delta_N\right)v_N^iv_{Nj}\rangle+o(\epsilon)\,.
\end{eqnarray}
Simplifying further using the average of \eqref{eq:ddotpsi-identity} and then inverting the Laplacian,
we obtain
\begin{eqnarray}\label{eq:EiiL}
  &&2\ddot{X}+6\frac{\dot{a}}{a}\dot{X}-4\frac{\dot{a}}{a}\dot{\overline{X}}+\left[4\partial_\tau\left(\frac{\dot{a}}{a}\right)+2\left(\frac{\dot{a}}{a}\right)^2\right]\left(X-\overline{X}\right)\nonumber\\
  &=&-2\langle\ddot{\chi}\rangle - 6\frac{\dot{a}}{a}\langle\dot{\chi}\rangle-2\frac{\dot{a}}{a}\langle\dot{\xi}\rangle-\left[4\partial_\tau\left(\frac{\dot{a}}{a}\right)+2\left(\frac{\dot{a}}{a}\right)^2\right]\left(\langle\chi\rangle+\langle\xi\rangle\right)\nonumber\\
  &&-\frac{1}{3}\overline{\partial^i\psi_N\partial_i\psi_N} + \frac{8\pi\rho_0a^2}{3}\overline{(1+\delta_N)v_N^iv_{Ni}}+o(\epsilon)\,.
\end{eqnarray}
Here, the constant of integration was determined by requiring consistency
with \eqref{eq:Eiismooth}.  Thus, the trace of \eqref{eqn:Eij} is satisfied to 
$O(\epsilon)$ at long wavelengths provided that $X$ satisfies this second order
ordinary differential equation in time. The ``scalar parts'' of the long wavelength part of
\eqref{eqn:Eij} have now been satisfied to $O(\epsilon)$

Using \eqref{eq:BN-identity}, it is not
difficult to see that the long wavelength part of the 
divergence of  \eqref{eqn:Eij} is
solved to $O(\epsilon)$ at large scales, 
without any need for further corrections.
Thus, the ``vector part'' of \eqref{eqn:Eij} has been 
satisfied to the desired order at long wavelengths.  
Only the ``tensor part'' of \eqref{eqn:Eij} remains.
Substituting the dictionary and applying
the averaging operator $\langle\cdot\rangle$, we obtain
\begin{eqnarray}\label{eq:J}
  &&\frac{1}{a^2}\left\{\ddot{J}^i_{\phantom{i}j}+\frac{\dot{a}}{a}\dot{J}^i_{\phantom{i}{j}}-\partial^k\partial_kJ^i_{\phantom{i}j}\right\}\nonumber\\
  &=&-\frac{1}{a^2}\left\{\langle\ddot{\jmath}^i_{\phantom{i}j}\rangle+\frac{\dot{a}}{a}\langle\dot{\jmath}^i_{\phantom{i}{j}}\rangle-\partial^k\partial_k\langle \jmath^i_{\phantom{i}j}\rangle\right\}+\frac{2}{a^2}\left.\langle\partial^i\psi_N\partial_j\psi_N\rangle\right|_{\text{t}}+\left.8\pi\rho_0\langle(1+\delta_N)v_N^iv_{Nj}\rangle\right|_{\text{t}}+o(\epsilon)\nonumber\\
  &=&-\frac{1}{a^2}\left\{\langle\ddot{\jmath}^i_{\phantom{i}j}\rangle+\frac{\dot{a}}{a}\langle\dot{\jmath}^i_{\phantom{i}{j}}\rangle\right\}+\frac{2}{a^2}\left.\overline{\partial^i\psi_N\partial_j\psi_N}\right|_{\text{t}}+\left.8\pi\rho_0\overline{(1+\delta_N)v_N^iv_{Nj}}\right|_{\text{t}}+o(\epsilon)\,,
\end{eqnarray}
where we used the
average of \eqref{eq:h1} in the second line.  Thus the tensor part of \eqref{eqn:Eij} is
solved to $O(\epsilon)$ at large scales provided that 
$J_{ij}$ solves this wave equation. This completes the solution of the long wavelength
part of \eqref{eqn:Eij} to $O(\epsilon)$.

Next, we consider \eqref{eqn:E0i}.
Substituting from the dictionary, applying $\langle\cdot\rangle$, and
taking the divergence, we obtain
\begin{eqnarray}
  &&-\frac{2}{a^2}\partial^i\partial_i\left\{\langle\dot{\psi}_N\rangle+\dot{X}+\langle\dot{\chi}\rangle+\frac{\dot{a}}{a}\left(\langle\psi_N\rangle+X + \langle\chi\rangle+\langle\xi\rangle\right)\right\}\nonumber\\
  &=&8\pi\rho_0\partial^i\left((1+\delta_N)v_{Ni}+P_i\right)+o(\epsilon)\,.
\end{eqnarray}
Using the average of \eqref{eq:LdotpsiN} to simplify this expression,
we obtain
\begin{equation}\label{eq:E0iL}
  -\frac{2}{a^2}\partial^i\partial_i\left\{\dot{X}+\langle\dot{\chi}\rangle+\frac{\dot{a}}{a}\left(X+\langle\chi\rangle+\langle\xi\rangle\right)\right\}=8\pi\rho_0\partial^iP_i+o(\epsilon)\,.
\end{equation}
We solve this equation by setting $P^i$ to be
\begin{equation}\label{eq:Pisdef}
P^{i}=-\frac{1}{4\pi\rho_0a^2}\partial^i\left(\dot{X}+\langle\dot{\chi}\rangle+\frac{\dot{a}}{a}\left(X+\langle\chi\rangle+\langle\xi\rangle\right)\right)\,.
\end{equation}
This satisfies the ``scalar part'' of \eqref{eqn:E0i} to the desired
order at long wavelengths.
It is easy to check that the vector part of \eqref{eqn:E0i} is also
satisfied without the need for any further corrections.

Finally, we consider the remaining component of Einstein's equation, the
time-time component \eqref{eqn:E00}. Substituting and averaging, we find
that this equation is satisfied to the required order by making the
density correction
\begin{eqnarray}\label{eq:Delta}
  \Delta&=&-\langle(1+\delta_N)v_N^iv_{Ni}\rangle-\frac{5}{8\pi a^2\rho_0}\langle\partial^i\psi_N\partial_i\psi_N\rangle-\frac{3}{8\pi\rho_0a^2}\left\{-\frac{2}{3}\partial^i\partial_i\left(X+\langle\chi\rangle\right)\right.\nonumber\\
  &&\left.+2\frac{\dot{a}}{a}\left(\dot{X}+\langle\dot{\chi}\rangle\right)+2\left(\frac{\dot{a}}{a}\right)^2\left(X-2\overline{X}+\langle\chi\rangle+\langle\xi\rangle\right)\right\}\nonumber\\
  &=&-\overline{(1+\delta_N)v_N^iv_{Ni}}-\frac{5}{8\pi a^2\rho_0}\overline{\partial^i\psi_N\partial_i\psi_N}-\frac{3}{8\pi\rho_0a^2}\left\{-\frac{2}{3}\partial^i\partial_iX\right.\nonumber\\
  &&\left.+2\frac{\dot{a}}{a}\left(\dot{X}+\langle\dot{\chi}\rangle\right)+2\left(\frac{\dot{a}}{a}\right)^2\left(X-2\overline{X}+\langle\chi\rangle+\langle\xi\rangle\right)\right\}\,.
\end{eqnarray}
Here, the average of \eqref{eq:chi1} was used to get the second line.

Einstein's equation has now been fully solved to $O(1)$ everywhere,
and to $O(\epsilon)$ at large scales. All of the quantities appearing
in our dictionary are uniquely determined by the Newtonian solution,
except for $X$ and $J_{ij}$, which obey second order differential
equations in time. The degrees of freedom associated with $X$
correspond to the long wavelength degrees of freedom present in the
dust matter sector in ordinary linearized perturbation theory.  It
would be natural to fix $X$ by requiring that $\Delta$ and $P_i$
vanish at an initial time\footnote{Since $\overline{P}_i$ vanishes identically,
this does not fix the spatially homogeneous part, $\overline{X}$, of
$X$. An additional condition on $\overline{X}$ will be imposed
in Appendix \ref{sec:modifiedbackground} .}.  The
degrees of freedom associated with $J_{ij}$ correspond to the presence
of long wavelength gravitational radiation. 

Finally, we consider the magnitude of the additional long wavelength
quantities $\Xi=-2\overline{X}$, $X$, $P_i$, $\Delta$, and $J_{ij}$
that we have just obtained. The equations for these 
quantities involve
terms of the form $\langle \partial_i \psi_N \partial_j \psi_N \rangle$
and $\langle(1+\delta_N)v_{Ni}v_{Nj}\rangle$ as well as
$\langle \xi \rangle$, $\langle \chi \rangle$, and 
$\langle \jmath_{ij}\rangle$,
which themselves are sourced by terms of the form
$\langle \partial_i \psi_N \partial_j \psi_N \rangle$
and $\langle(1+\delta_N)v_{Ni}v_{Nj}\rangle$.
Thus, the additional
long wavelength quantities appearing in our new dictionary
\eqref{eqn:d1A}--\eqref{eqn:d5A} 
should have a magnitude of the order of the Newtonian
potential energy and kinetic energy of the dust matter. 
Although very small, the homogeneous (i.e., spatially constant) part of 
these quantities provides the dominant correction to the background
FLRW dust cosmology. We compute these corrections explicitly in 
Appendix \ref{sec:modifiedbackground}. However, the long wavelength
corrections at finite wavelength are sourced by large scale inhomogeneities
in $\langle \partial_i \psi_N \partial_j \psi_N \rangle$
and $\langle(1+\delta_N)v_{Ni}v_{Nj}\rangle$. These 
terms should be {\em extremely}
small as compared with, say, $\langle \delta_N \rangle$. Thus,
the long wavelength corrections we have obtained in this subsection
should make entirely negligible contributions to 
Newtonian large scale structure.

\section{Summary}\label{sec:summary}

Combining all of the results of the previous section, we have the following
{\em Oxford dictionary} for translating 
a Newtonian cosmological solution
$(\psi_N,v_N^i,\delta_N)$ to a general relativistic spacetime metric
\eqref{eqn:Poissongauge} and dust stress-energy \eqref{Tab}:
\begin{eqnarray}
  \label{eqn:d1B}A&=&\psi_N+\chi+\xi+X-2\overline{X}\,,\\
  \label{eqn:d2B}H_L&=&-\psi_N-\chi-X\,,\\
  \label{eqn:d3B}(1+\delta_N)v_i&=&(1+\delta_N)(v_{Ni}+ B_i)-\left.\overline{(1+\delta_N)v_{Ni}}\right|_{\text{v}}+P_i\,,\\
  \label{eqn:d4B}\delta&=&\delta_N-\frac{3}{4\pi\rho_0a^2}\left[\left(\frac{\dot{a}}{a}\right)^2\psi_N+\frac{\dot{a}}{a}\dot{\psi}_N\right]+\Delta\,,\\
  \label{eqn:d5B}h_{ij}&=&\jmath_{ij}+J_{ij}\,.
\end{eqnarray}
Here, the quantities $B_i$, $\xi$, $\chi$, $\jmath_{ij}$, $P^i$, and $\Delta$ 
are given by
{\allowdisplaybreaks\begin{eqnarray}
  \label{eqn:d6B}\partial^j\partial_j B^i&=&-16\pi\rho_0a^2\left.\left((1+\delta_N)v_N^i-\overline{(1+\delta_N)v_N^i}\right)\right|_{\text{v}}\,,\\
  \label{eqn:d7B}\partial^i\partial_i\partial^j\partial_j\xi&=&3\partial_i\partial^j\left\{2\psi_N\partial^i\partial_j\psi_N+\partial^i\psi_N\partial_j\psi_N\right\}-\partial^i\partial_i\left\{2\psi_N\partial^j\partial_j\psi_N+\partial^j\psi_N\partial_j\psi_N\right\}\nonumber\\*
  &&-12\pi\rho_0a^2\partial_i\partial^j\left[(1+\delta_N)v_N^iv_{Nj}\right]+4\pi\rho_0a^2\partial^i\partial_i\left[(1+\delta_N)v_N^kv_{Nk}\right]\,,\\
  \label{eqn:d8B}\partial^i\partial_i\chi&=&-4\psi_N\partial^i\partial_i\psi_N-\frac{3}{2}\partial^i\psi_N\partial_i\psi_N+4\pi a^2\rho_0(1+\delta_N)v_N^iv_{Ni}\nonumber\\*
  &&-\frac{5}{2}\overline{\partial^i\psi_N\partial_i\psi_N}-4\pi a^2\rho_0\overline{(1+\delta_N)v_N^iv_{Ni}}\,,\\
  \label{eqn:d9B}\partial^k\partial_k\jmath^i_{\phantom{i}j}&=&\left.\left\{8\pi\rho_0a^2(1+\delta_N)v_N^iv_{Nj}+4\psi_N\partial^i\partial_j\psi_N+2\partial^i\psi_N\partial_j\psi_N\right.\right.\nonumber\\*
  &&\left.\left.-8\pi\rho_0a^2\overline{(1+\delta_N)v_N^iv_{Nj}}+2\overline{\partial^i\psi_N\partial_j\psi_N}\right\}\right|_{\text{t}}\,,\\
  \label{eqn:d11B}P^{i}&=&-\frac{1}{4\pi\rho_0a^2}\partial^i\left(\dot{X}+\langle\dot{\chi}\rangle+\frac{\dot{a}}{a}\left(X+\langle\chi\rangle+\langle\xi\rangle\right)\right)\,,\\
  \label{eqn:d12B}\Delta&=&-\overline{(1+\delta_N)v_N^iv_{Ni}}-\frac{5}{8\pi a^2\rho_0}\overline{\partial^i\psi_N\partial_i\psi_N}-\frac{3}{8\pi\rho_0a^2}\left\{-\frac{2}{3}\partial^i\partial_iX\right.\nonumber\\*
  &&\left.+2\frac{\dot{a}}{a}\left(\dot{X}+\langle\dot{\chi}\rangle\right)+2\left(\frac{\dot{a}}{a}\right)^2\left(X-2\overline{X}+\langle\chi\rangle+\langle\xi\rangle\right)\right\}\,,
\end{eqnarray}}
with $\overline{B}_i = \overline{\xi} = \overline{\chi} = \overline{\jmath}_{ij} =0$.
The quantities $J_{ij}$ and $X$ satisfy the differential equations
\begin{eqnarray}\label{eqn:d13B}
  &&\frac{1}{a^2}\left\{\ddot{J}^i_{\phantom{i}j}+\frac{\dot{a}}{a}\dot{J}^i_{\phantom{i}{j}}-\partial^k\partial_kJ^i_{\phantom{i}j}\right\}\nonumber\\
  &=&-\frac{1}{a^2}\left\{\langle\ddot{\jmath}^i_{\phantom{i}j}\rangle+\frac{\dot{a}}{a}\langle\dot{\jmath}^i_{\phantom{i}{j}}\rangle\right\}+\frac{2}{a^2}\left.\overline{\partial^i\psi_N\partial_j\psi_N}\right|_{\text{t}}+\left.8\pi\rho_0\overline{(1+\delta_N)v_N^iv_{Nj}}\right|_{\text{t}}\,.
\end{eqnarray}
and
\begin{eqnarray}\label{eqn:d14B}
  &&2\ddot{X}+6\frac{\dot{a}}{a}\dot{X}-4\frac{\dot{a}}{a}\dot{\overline{X}}+\left[4\partial_\tau\left(\frac{\dot{a}}{a}\right)+2\left(\frac{\dot{a}}{a}\right)^2\right]\left(X-\overline{X}\right)\nonumber\\
  &=&-2\langle\ddot{\chi}\rangle - 6\frac{\dot{a}}{a}\langle\dot{\chi}\rangle-2\frac{\dot{a}}{a}\langle\dot{\xi}\rangle-\left[4\partial_\tau\left(\frac{\dot{a}}{a}\right)+2\left(\frac{\dot{a}}{a}\right)^2\right]\left(\langle\chi\rangle+\langle\xi\rangle\right)\nonumber\\
  &&-\frac{1}{3}\overline{\partial^i\psi_N\partial_i\psi_N} + \frac{8\pi\rho_0a^2}{3}\overline{(1+\delta_N)v_N^iv_{Ni}}\,.
\end{eqnarray}
Like the Oxford English Dictionary, the above dictionary should be far
more detailed and precise than needed for everyday use. Nevertheless,
it may be comforting to have it on one's bookshelf in case the need
does arise. Furthermore, as a matter of principle, it is of
importance to know that a dictionary of this accuracy---namely,
solving Einstein's equation to $O(1)$ on all scales and to
$O(\epsilon)$ on large scales---can be constructed without running
into inconsistencies.

Our main purpose in obtaining the complete dictionary
\eqref{eqn:d1B}--\eqref{eqn:d5B} was to evaluate the accuracy of the
original dictionary \eqref{eq:dict-L1}--\eqref{eq:dict-L5} (as well as
its abridgement \eqref{eq:adict-L1}--\eqref{eq:adict-L3} and
simplification \eqref{eq:sdict-L1}--\eqref{eq:sdict-L3}).  We have
argued that for a Newtonian cosmology that satisfies $\psi_N \ll 1$
and $|v_N^i| \ll 1$ but may have $\delta_N \gg 1$ at small scales,
all of the additional terms appearing in
\eqref{eqn:d1B}--\eqref{eqn:d5B} as compared with
\eqref{eq:dict-L1}--\eqref{eq:dict-L5} should be negligibly
small. Whether or not this is actually the case for any given
Newtonian cosmology can be determined by computing the quantities
$\xi$, $\chi$, $\jmath_{ij}$, $P^i$, $\Delta$, $X$, and $J_{ij}$ given
by \eqref{eqn:d7B}--\eqref{eqn:d14B}. If these quantities are indeed
negligibly small, then one can have confidence that the Newtonian
cosmology is accurately representing a general relativistic spacetime
via the original dictionary
\eqref{eq:dict-L1}--\eqref{eq:dict-L5}. If, in addition, $B^i$ is
negligibly small (see \eqref{eqn:d6B}), then one is similarly
justified in using the abridged dictionary
\eqref{eq:adict-L1}--\eqref{eq:adict-L3}. These statements remain
valid even in cases where the Newtonian cosmology is describing
phenomena on scales comparable to or larger than the Hubble radius.

\begin{acknowledgments}

  We thank Todd Oliynyk for helpful discussions.  This research was
  supported in part by NSF Grant No.~PHY08-54807 to the University of
  Chicago.

\end{acknowledgments}

\appendix

\section{Einstein's equation}\label{sec:Einstein}

For the metric ansatz \eqref{eqn:Poissongauge}, we write down the
various components of the perturbed Einstein equation, 
\begin{equation}
  G^\mu_{\phantom{\mu}\nu}(g)-G^\mu_{\phantom{\mu}\nu}(g^{(0)})=8\pi\left(T^\mu_{\phantom{\mu}\nu}-T^{(0)}{}^\mu_{\phantom{\mu}\nu}\right)\,,
\end{equation}
keeping all terms which are $O(1)$ at small scales or $O(\epsilon)$ at
large scales in our counting scheme.  We introduce the notation
$o(1;\epsilon)$ to denote a quantity which is $o(1)$ at small scales
and $o(\epsilon)$ at large scales.  The
${}^\mu_{\phantom{\mu}\nu}={}^0_{\phantom{0}0}$ equation reads
\begin{eqnarray}\label{eqn:E00}
  &&\frac{3}{a^2}\left\{-2\frac{\dot{a}}{a}\dot{H}_L+\frac{2}{3}\partial^i\partial_iH_L+2\left(\frac{\dot{a}}{a}\right)^2A-\frac{8}{3}H_L\partial^i\partial_iH_L-\partial_iH_L\partial^iH_L\right\}\nonumber\\ 
  &=&8\pi\rho_0\left\{-\delta-(1+\delta)v^i(v_i-B_i)\right\}+o(1;\epsilon)\,,
\end{eqnarray}
the ${}^\mu_{\phantom{\mu}\nu}={}^0_{\phantom{0}i}$ equation is
\begin{equation}\label{eqn:E0i}
  \frac{2}{a^2}\left\{\partial_i\dot{H}_L-\frac{\dot{a}}{a}\partial_iA\right\}-\frac{1}{2a^2}\partial^j\partial_jB_i=8\pi\rho_0(1+\delta)(v_i-B_i)+o(1;\epsilon)\,,
\end{equation}
and the ${}^\mu_{\phantom{\mu}\nu}={}^i_{\phantom{0}j}$ equation is
\begin{eqnarray}\label{eqn:Eij}
&&\frac{1}{a^2}\left\{\partial^k\partial_k(H_L+A)-2\ddot{H}_L-4\frac{\dot{a}}{a}\dot{H}_L+2\frac{\dot{a}}{a}\dot{A}+4\partial_\tau\left(\frac{\dot{a}}{a}\right)A+2\left(\frac{\dot{a}}{a}\right)^2A\right\}\delta^i_{\phantom{i}j}\nonumber\\
&&+\frac{1}{a^2}\left\{-4H_L\partial^k\partial_kH_L-2A\partial^k\partial_kA-2H_L\partial^k\partial_kA-2\partial_kH_L\partial^kH_L-\partial_kA\partial^kA\right\}\delta^i_{\phantom{i}j}\nonumber\\
&&-\frac{1}{a^2}\partial^i\partial_j(H_L+A)+\frac{1}{a^2}\left\{4H_L\partial^i\partial_jH_L+2H_L\partial^i\partial_jA+2A\partial^i\partial_jA+\partial^iA\partial_jA\right.\nonumber\\
&&\left.+\partial^iA\partial_jH_L+\partial^iH_L\partial_jA+3\partial^iH_L\partial_jH_L\right\}\nonumber\\
&&+\frac{1}{2a^2}\left\{\partial^i\dot{B}_j+\partial_j\dot{B}^i+2\frac{\dot{a}}{a}\partial^iB_j+2\frac{\dot{a}}{a}\partial_jB^i\right\}+\frac{1}{a^2}\left\{\ddot{h}^i_{\phantom{i}j}+2\frac{\dot{a}}{a}\dot{h}^i_{\phantom{i}j}-\partial^k\partial_kh^i_{\phantom{i}j}\right\}\nonumber\\
&=&8\pi\rho_0(1+\delta)v^i(v_j-B_j)+o(1;\epsilon)\,.
\end{eqnarray}

\section{Modified background metric}\label{sec:modifiedbackground}

In this Appendix we compute the homogeneous part of the metric and
matter distribution as given by our final dictionary
\eqref{eqn:d1B}--\eqref{eqn:d5B}.  These can be viewed as providing the
dominant corrections to the background cosmology produced by small
scale inhomogeneities. The relevant equations can be obtained by
taking spatial integrals of the equations of section
\ref{sec:summary}.  We find that the spatially homogeneous parts of
the metric components are given by
\begin{eqnarray}
 \overline{A}&=&-\overline{X}\,,\\
 \overline{H}_L&=&-\overline{X} \,,\\
 \overline{h}_{ij}&=&\overline{J}_{ij}\,,
\end{eqnarray}
as well as $\overline{B}^i = 0$ (see footnote \ref{Bbar}).
Thus, the homogeneous part of the metric takes the form
\begin{equation}
  \overline{ds}^2=a^2(\tau)\left[-(1-2\overline{X})d\tau^2 + ((1-2\overline{X})\delta_{ij}+ \overline{J}_{ij}) dx^idx^j\right]\,.
\end{equation}
We also have
\begin{equation}
 \overline{\delta}=\overline{\Delta} \,,
\end{equation}
\begin{equation}
\overline{P}^i = 0\,,
\end{equation}
and
\begin{equation}
\overline{(1+\delta)(v_i - B_i)}= 0\,.
\end{equation}
In addition, the quantities $\overline{X}$, $\overline{\Delta}$, and $\overline{J}_{ij}$ satisfy
\begin{equation}
  \label{eq:barijA}2\ddot{\overline{X}}+2\frac{\dot{a}}{a}\dot{\overline{X}}-\left[4\partial_\tau\left(\frac{\dot{a}}{a}\right)+2\left(\frac{\dot{a}}{a}\right)^2\right]\overline{X}=-\frac{1}{3}\overline{\partial^i\psi_N\partial_i\psi_N} + \frac{8\pi}{3}a^2\rho_0 \overline{ (1+\delta_N) v_N^iv_{Ni}}\,,
\end{equation}
\begin{equation}
  \label{eq:bar00A}\overline{\Delta}=-\overline{(1+\delta_N)v_N^iv_{Ni}}-\frac{5}{8\pi a^2\rho_0}\overline{\partial^i\psi_N\partial_i\psi_N}-\frac{3}{8\pi\rho_0a^2}\left\{2\frac{\dot{a}}{a}\dot{\overline{X}}-2\left(\frac{\dot{a}}{a}\right)^2\overline{X}\right\}\,,
\end{equation}
and
\begin{equation}\label{eq:hbar}
  \frac{1}{a^2}\left\{\ddot{\overline{J}}^i_{\phantom{i}j}+\frac{\dot{a}}{a}\dot{\overline{J}}^i_{\phantom{i}{j}}\right\}=\left.\left\{\frac{2}{a^2}\overline{\partial^i\psi_N\partial_j\psi_N}+8\pi\rho_0\overline{(1+\delta_N)v_N^iv_{Nj}}\right\}\right|_{\text{t}}\,.
\end{equation}

It is clear that the metric perturbation given by $\overline{X}$ can be
interpreted as taking one to a new FLRW spacetime, with scale factor
\begin{equation}
  \hat{a}(\tau)=a(\tau)(1-\overline{X})\,.
\end{equation}
We now derive modified Friedmann equations for $\hat{a}$.  To linear
order in barred quantities, we have
\begin{equation}
  \frac{1}{\hat{a}}\frac{d\hat{a}}{d\tau}=\frac{1}{a}\frac{da}{d\tau}-\dot{\overline{X}}\,,
\end{equation}
so
\begin{eqnarray}\label{eq:newFriedmann1}
  \left(\frac{1}{\hat{a}}\frac{d\hat{a}}{d\tau}\right)^2&=&\left(\frac{1}{a}\frac{da}{d\tau}\right)^2-2\frac{\dot{a}}{a}\dot{\overline{X}}\nonumber\\
  &=&\frac{8\pi\rho_0\hat{a}^2}{3}\left(1+\overline{\Delta}+\overline{(1+\delta_N)v_N^iv_{Ni}}\right)+\frac{5}{3}\overline{\partial^i\psi_N\partial_i\psi_N}+\frac{\Lambda \hat{a}^2}{3}\,.
\end{eqnarray}
Here, we made use of the Friedmann equation for $a$
as well as eq.~\eqref{eq:bar00A}.  Similarly, we have
\begin{eqnarray}\label{eq:newFriedmann2}
  \frac{d}{d\tau}\left(\frac{1}{\hat{a}}\frac{d\hat{a}}{d\tau}\right)&=&\frac{d}{d\tau}\left(\frac{1}{a}\frac{da}{d\tau}\right)-\ddot{\overline{X}}\nonumber\\
  &=&-\frac{4\pi\rho_0\hat{a}^2}{3}\left(1+\overline{\Delta}+2\overline{(1+\delta_N)v_N^iv_{Ni}}\right)-\frac{2}{3}\overline{\partial^i\psi_N\partial_i\psi_N}+\frac{\Lambda\hat{a}^2}{3}\,.
\end{eqnarray}

To put these equations in a more recognizable form, we note that for dust matter, 
 $\nabla_a (\rho u^a) = 0$, so the integrated flux of $\rho u^a$ 
over a Cauchy surface $\Sigma$ 
\begin{equation}
{\mathcal N} = -\int_\Sigma{\rho u^d \epsilon_{dabc}}
\label{N}
\end{equation}
is a constant, i.e., independent of $\Sigma$. ${\mathcal N}$ is often referred to as the
``total number of baryons''; in an $N$-body simulation, it would correspond to the total
number of particles in the simulation.
Evaluating the right side of \eqref{N}, we obtain
\begin{equation}
{\mathcal N} = \rho_0 a^3 \left(1+\overline{\Delta}+\frac{1}{2}\overline{(1+\delta_N)v_N^iv_{Ni}}+\frac{3}{4\pi\rho_0 a^2} \overline{\partial^i\psi_N\partial_i\psi_N} - 3\overline{X}\right)\,.
\end{equation}
It is natural to use our freedom in choosing initial conditions for a solution to \eqref{eq:barijA} to require
${\mathcal N} = {\mathcal N}_0 = \rho_0 a^3$, so that the total number of particles is the same
as in the background spacetime. This condition yields
\begin{equation}
  \overline{\Delta}=-\frac{1}{2}\overline{(1+\delta_N)v_N^iv_{Ni}}-\frac{3}{4\pi\rho_0 a^2} \overline{\partial^i\psi_N\partial_i\psi_N}+3\overline{X}\,.
\end{equation}
Combining this equation with \eqref{eq:bar00A}, we obtain
\begin{equation}\label{eq:bar00B}
  0=-\frac{1}{2}\overline{(1+\delta_N)v_N^iv_{Ni}}+\frac{1}{8\pi a^2\rho_0}\overline{\partial^i\psi_N\partial_i\psi_N}-3\overline{X}-\frac{3}{8\pi a^2\rho_0}\left\{2\frac{\dot{a}}{a}\dot{\overline{X}}-2\left(\frac{\dot{a}}{a}\right)^2\overline{X}\right\}\,,
\end{equation}
whose time derivative is \eqref{eq:barijA}.

We define the average particle number density $\hat{\rho}$ relative to $\hat{a}$ by
\begin{equation}
\hat{\rho}\hat{a}^3 = {\mathcal N}= {\mathcal N}_0 = \rho_0 a^3\,.
\end{equation}
In terms of $\hat{\rho}$, the Friedmann equations become
\begin{eqnarray}
  \left(\frac{1}{\hat{a}}\frac{d\hat{a}}{d\tau}\right)^2&=&\frac{8\pi\hat{\rho}\hat{a}^2}{3}\left(1+\frac{1}{2}\overline{(1+\delta_N)v_N^iv_{Ni}}-\frac{1}{8\pi\hat{\rho}\hat{a}^2}\overline{\partial^i\psi_N\partial_i\psi_N}\right)+\frac{\Lambda \hat{a}^2}{3}\,,\\
  \frac{d}{d\tau}\left(\frac{1}{\hat{a}}\frac{d\hat{a}}{d\tau}\right)&=&-\frac{4\pi\hat{\rho}\hat{a}^2}{3}\left(1+\frac{3}{2}\overline{(1+\delta_N)v_N^iv_{Ni}}-\frac{1}{4\pi\hat{\rho}\hat{a}^2}\overline{\partial^i\psi_N\partial_i\psi_N}\right)+\frac{\Lambda\hat{a}^2}{3}\,.
\end{eqnarray}
From these equations, one can read off the effective energy density
and pressure, including the contributions from small scale inhomogeneities,
\begin{eqnarray}
  \label{eq:rhoeff}\rho_{\text{eff}}&=&\hat{\rho}\left(1+\frac{1}{2}\overline{(1+\delta_N)v_N^iv_{Ni}}-\frac{1}{8\pi\hat{\rho}\hat{a}^2}\overline{\partial^i\psi_N\partial_i\psi_N}\right)\,,\\
  \label{eq:Peff}P_{\text{eff}}&=&\hat{\rho}\left(\frac{1}{3}\overline{(1+\delta_N)v_N^iv_{Ni}}-\frac{1}{24\pi\hat{\rho}\hat{a}^2}\overline{\partial^i\psi_N\partial_i\psi_N}\right)\,.
\end{eqnarray}
The correction terms in $\rho_{\text{eff}}$ correspond precisely to
averaged gravitational potential energy and kinetic energy, as
expected \cite{Futamase:1996fk,Baumann:2010tm,Green:2010qy}.  They can
be interpreted as renormalizing the proper mass density $\hat{\rho}$
to an ``ADM mass density'' $\rho_{\text{eff}}$.  For virialized
systems, the correction terms in $P_{\text{eff}}$ cancel, as pointed
out in \cite{Baumann:2010tm}.  Thus, we see that the corrections
resulting from $\overline{X}$ and $\overline{\Delta}$ correspond to
modifying the FLRW background to a new FLRW spacetime with small
corrections to the average effective mass density and pressure that
arise from small scale Newtonian gravitational potential energy and
stresses as well as small scale kinetic motions.

The remaining corrections due to $\overline{J_{ij}}$ perturb one to an
anisotropically expanding Bianchi model. It can be seen from
\eqref{eq:hbar} that anisotropies in the spatial average of the
Newtonian stresses and/or kinetic motions must necessarily induce an
anisotropic expansion of the universe. However, we would expect these
effects to be extremely small.

\bibliography{mybib.bib}

\end{document}